\documentclass[sn-nature]{sn-jnl}

\usepackage{graphicx}%
\usepackage{multirow}%
\usepackage{amsmath,amssymb,amsfonts}%
\usepackage{amsthm}%
\usepackage{mathrsfs}%
\usepackage[title]{appendix}%
\usepackage{xcolor}%
\usepackage{textcomp}%
\usepackage{manyfoot}%
\usepackage{booktabs}%
\usepackage{algorithm}%
\usepackage{algorithmicx}%
\usepackage{algpseudocode}%
\usepackage{listings}%

\usepackage{placeins}  

\usepackage{rotating}
\usepackage{tikz}

\begin{document}

\title[Global spatio-temporal downscaling of ERA5 precipitation through generative AI]{Global spatio-temporal downscaling of ERA5 precipitation through generative AI}


\author*[1]{\fnm{Luca} \sur{Glawion}}\email{luca.glawion@kit.edu}

\author[1,2]{\fnm{Julius} \sur{Polz}}\email{julius.polz@kit.edu}
\author[1,3]{\fnm{Harald} \sur{Kunstmann}}\email{harald.kunstmann@kit.edu}
\author[1]{\fnm{Benjamin} \sur{Fersch}}\email{benjamin.fersch@kit.edu}
\author[1]{\fnm{Christian} \sur{Chwala}}\email{christian.chwala@kit.edu}

\affil[1]{\orgdiv{Institute of Meteorology and Climate Research - Atmospheric Environmental Research (IMK-IFU), Campus Alpin}, \orgname{Karlsruhe Institute of Technology}, \orgaddress{\city{Garmisch-Partenkirchen}, \country{Germany}}}

\affil[2]{\orgdiv{Institute of Meteorology and Climate Research - Atmospheric Trace Gases and Remote Sensing (IMK-ASF)}, \orgname{Karlsruhe Institute of Technology}, \orgaddress{\city{Karlsruhe}, \country{Germany}}}

\affil[3]{\orgdiv{Institute of Geography}, \orgname{University of Augsburg}, \orgaddress{\city{Augsburg}, \country{Germany}}}

\abstract{The spatial and temporal distribution of precipitation has a significant impact on human lives by determining freshwater resources and agricultural yield, but also rainfall-driven hazards like flooding or landslides. While the ERA5 reanalysis dataset provides consistent long-term global precipitation information that allows investigations of these impacts, it lacks the resolution to capture the high spatio-temporal variability of precipitation. ERA5 misses intense local rainfall events that are crucial drivers of devastating flooding - a critical limitation since extreme weather events become increasingly frequent.
\\
Here, we introduce spateGAN-ERA5, the first deep learning based spatio-temporal downscaling of precipitation data on a global scale. SpateGAN-ERA5 uses a conditional generative adversarial neural network (cGAN) that enhances the resolution of ERA5 precipitation data from 24\,km and 1 hour to 2\,km and 10 minutes, delivering high-resolution rainfall fields with realistic spatio-temporal patterns and accurate rain rate distribution including extremes. Its computational efficiency enables the generation of a large ensemble of solutions, addressing uncertainties inherent to the challenges of downscaling.
\\
Trained solely on data from Germany and validated in the US and Australia considering diverse climate zones, spateGAN-ERA5 demonstrates strong generalization indicating a robust global applicability. SpateGAN-ERA5 fulfils a critical need for high-resolution precipitation data in hydrological and meteorological research, offering new capabilities for flood risk assessment, AI-enhanced weather forecasting, and impact modelling to address climate-driven challenges worldwide.}




\maketitle

\section{Main}\label{sec1}

Variations in precipitation critically influence society and ecosystems, affecting water resources, agriculture, and flood risks \cite{gherardi_effect_2019, kotz_effect_2022, ray_climate_2015}. Climate change has already amplified precipitation variability, leading to more frequent and severe weather events \cite{zhang_anthropogenic_2024}. Understanding and mitigating the impacts of precipitation extremes requires accurate historical records in a spatial and temporal resolution that captures the high variability of rainfall \cite{berne_temporal_2004, ochoa-rodriguez_impact_2015, tamm_intensification_2023}. Observation-based rainfall products can only partially fulfil this requirement. Station networks have long records, but are not dense enough in most parts of the world \cite{kidd_so_2017} and thus lack spatial representativeness. In contrast, satellite rainfall products provide homogeneous spatial coverage but only have limited temporal coverage. In addition, they suffer from considerable errors due to their complex rainfall retrieval methods and exhibit spatial and temporal inhomogeneities \cite{guo_inter-comparison_2015, guo_early_2016, bai_accuracy_2018}.\\
\\
Assimilation of historical meteorological observations in first-principle-based physical simulations enables modelling of consistent, comprehensive and long records of atmospheric conditions \cite{hersbach_era5_2020}. In the last decade, such reanalyses have accelerated scientific research in hydrological modelling \cite{gebrechorkos_global-scale_2024, tarek_evaluation_2020}, flood prediction \cite{nearing_global_2024}, calculation of climate change related costs \cite{kotz_economic_2024, lenton_quantifying_2023}, or training data-driven weather forecasting models \cite{lam_graphcast_2023, pathak_fourcastnet_2022, lang_aifs_2024, kochkov_neural_2024}. However, existing global reanalyses still have significant limitations. The heterogeneous density of assimilated observations and the low spatio-temporal model resolution lead to uncertainties and biases \cite{hersbach_era5_2020, lavers_evaluation_2022}. In particular, the complex spatio-temporal structure of rainfall cannot be represented by the resolution of current reanalysis products, which leads to a significant underestimation of extreme values, which are crucial for impact analysis of severe weather events \cite{volosciuk_extreme_2015, seo_real-time_1998, sun_flood_2000,aleshina_link_2021,ben-bouallegue_rise_2023}. Running higher-resolution global reanalyses is currently not feasible due to the immense computational demand \cite{tabari_local_2016, zandler_evaluation_2019, bjarke_evaluating_2024}.\\
\\
Downscaling can be used to increase the spatial and temporal resolution of coarse-resolution global models, either dynamically, that is running a local-area high-resolution model, or by statistical post-processing. While dynamical downscaling is again limited by computational resources, statistical methods are computationally efficient and can be applied globally. However, traditional statistical approaches are not capable of generating realistic high-resolution rainfall fields with correct spatio-temporal patterns \cite{bytheway_evaluation_2023} and extreme values. Recently, advanced downscaling approaches leveraging deep neural networks have proven to be capable of this task. Successful applications have been shown for spatial and spatio-temporal superresolution \cite{glawion_spategan_2023, leinonen_stochastic_2021, serifi_spatio-temporal_2021}, and regional spatial downscaling \cite{harris_generative_2022, price_increasing_2022}. Nevertheless, a skilful global sub-hourly, km-scale downscaling of precipitation data has remained a challenging problem. \\
\\
Here, we present spateGAN-ERA5, a conditional generative adversarial network for robust deep learning-based spatio-temporal downscaling. Our model transforms hourly, 24\,km ($\sim$0.25°) resolved ERA5 precipitation estimates into rainfields that resemble weather radar observations at a resolution of 10 minutes and 2\,km. SpateGAN-ERA5 is trained on high-resolution quantitative precipitation estimates (QPE) from a gauge-adjusted and climatology corrected weather radar product in Germany and is evaluated across three climatically diverse regions on the globe. The model generalises well outside the training domain and enables computationally efficient global rainfall downscaling to a resolution that is fine enough to capture the spatio-temporal complexity of rainfall, especially for rainfall events with convective cells. It generates realistic extreme value distributions, spatial structures, and advection patterns, all in a well-calibrated ensemble that addresses the underdetermined nature of the downscaling problem. Thus, spateGAN-ERA5 significantly advances downscaling methodologies and opens up a wide field of possible scientific investigations in a variety of domains like hydrology, risk analysis or agriculture.

\section{Generative spatio-temporal downscaling of global ERA5 precipitation}\label{sec2}

\begin{figure}[ht]
\centering
\includegraphics[width=1\textwidth]{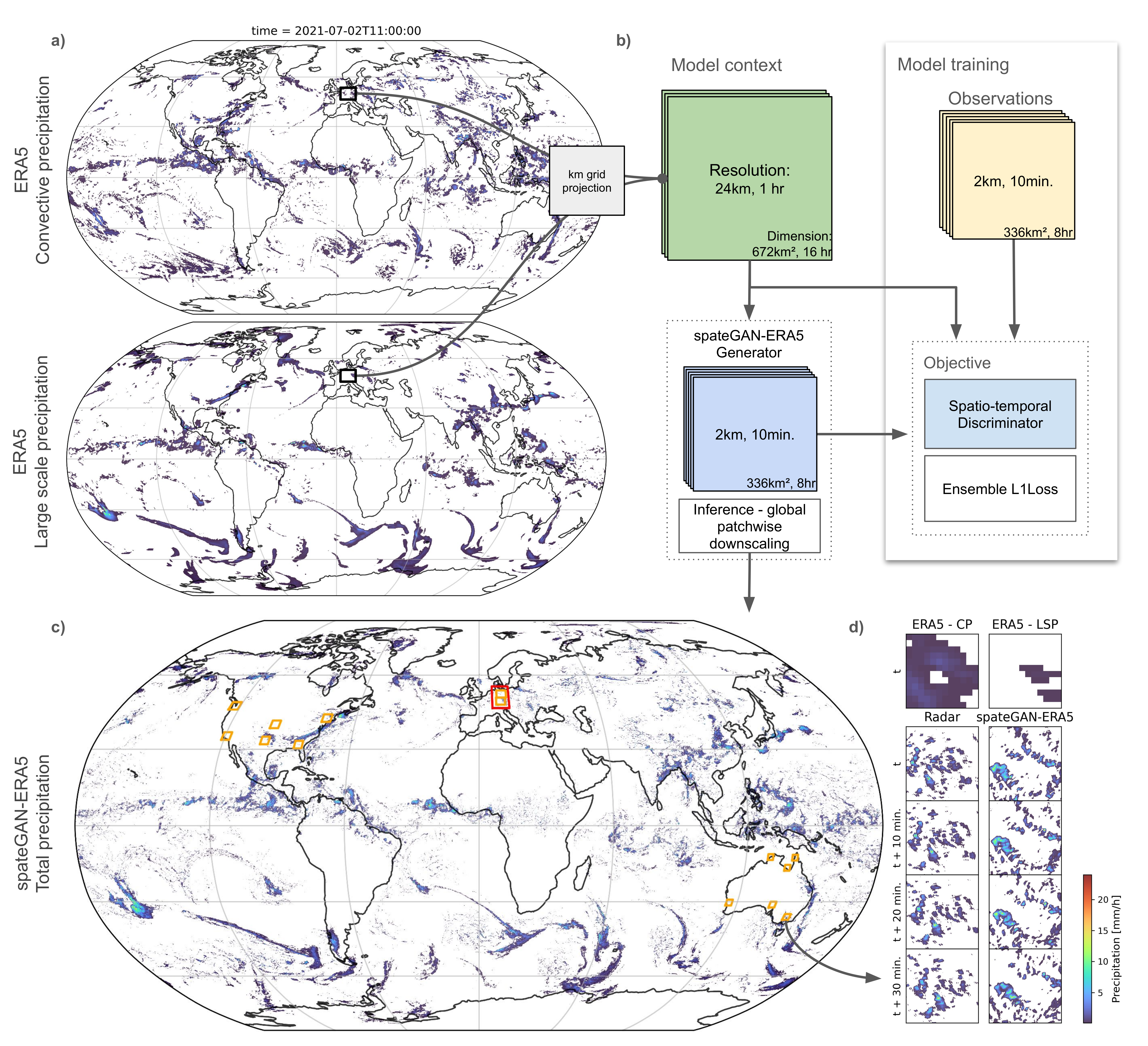}
\caption{Model and evaluation area overview for spatio-temporal downscaling of global ERA5 precipitation estimates. spateGAN-ERA5 can transform coarse-resolution ERA5 rainfields shown in a) into high-resolution rainfields (2\,km, 10\,min.) as they would be observed by a gauge-adjusted radar product regarding their spatial structures and rain rate distribution. b) Schematic of the model operating on patchwise downscaling of km gridded convective and large-scale ERA5 precipitation variables in a probabilistic manner. c) Global downscaling predictions enabled by patch stitching provide continuous rainfields (full resolution map shown in ancillary files
Fig. A1). From the area marked by the red box, patches are drawn and used for model training. Downscaling performance is evaluated using radar observations as a comparison from the regions marked by the orange boxes. d) Detailed highlight shows the resolved resolution in time and space and the comparison with the Australian weather radar observations.
}\label{fig1}
\end{figure}

The central methodology we present for global downscaling of ERA5 precipitation data is the use of a conditional generative adversarial network (cGAN) with ERA5 convective (CP) and large-scale precipitation (LSP) as the coarse condition and gauge-adjusted weather radar data as the high-resolution reference (see Fig. \ref{fig1}b). The downscaling of hourly ERA5 precipitation fields with a spatial resolution of 24\,km is performed by a generator model producing a field with a 12-times higher spatial and a 6-times higher temporal resolution. Specifically, the generator processes CP and LSP input patches with a size of 28 by 28 grid cells and 16 time steps. To provide more contextual information, the input is four times the domain size of the actual downscaled area (see Fig. \ref{fig2}a).\\
\\
A main feature of our cGAN is the custom learnable loss function (the discriminator) which enables the generation of realistic fields that fulfil a wide range of statistical and structural criteria for precipitation. The applied neural network architecture extends the spateGAN model established for a weather radar video-super-resolution approach \cite{glawion_spategan_2023} and is described in section \ref{sec5_1}. The model is trained using high-quality gauge-adjusted weather radar data provided by the German Meteorological Service (DWD) from the years 2009-2020 \cite{winterrath_radar_2018}. Details on the adversarial training procedure are given in section \ref{sec5_2}. The model is efficient, fast, and small enough to run on a single NVIDIA-Tesla-V100 GPU by downscaling one patch in 0.04 s in inference mode. Data-parallel training on 4 A100 GPUs took 3 days.\\
\\
Global fields are produced by stitching overlapping high-resolution patches (see section \ref{sec5_8}). We evaluate the downscaling skill by comparison to weather radar data from the year 2021 in three different countries (Germany, USA, Australia) that cover a wide range of climatic conditions (see Fig. \ref{fig1}c). Performance is compared to the rainFARM downscaling technique \cite{rebora_rainfarm_2006, donofrio_stochastic_2014} and to trilinear interpolation as a simple baseline method.

\section{Case Study}\label{sec3}

\begin{figure}[ht]
\centering
\includegraphics[width=1\textwidth]{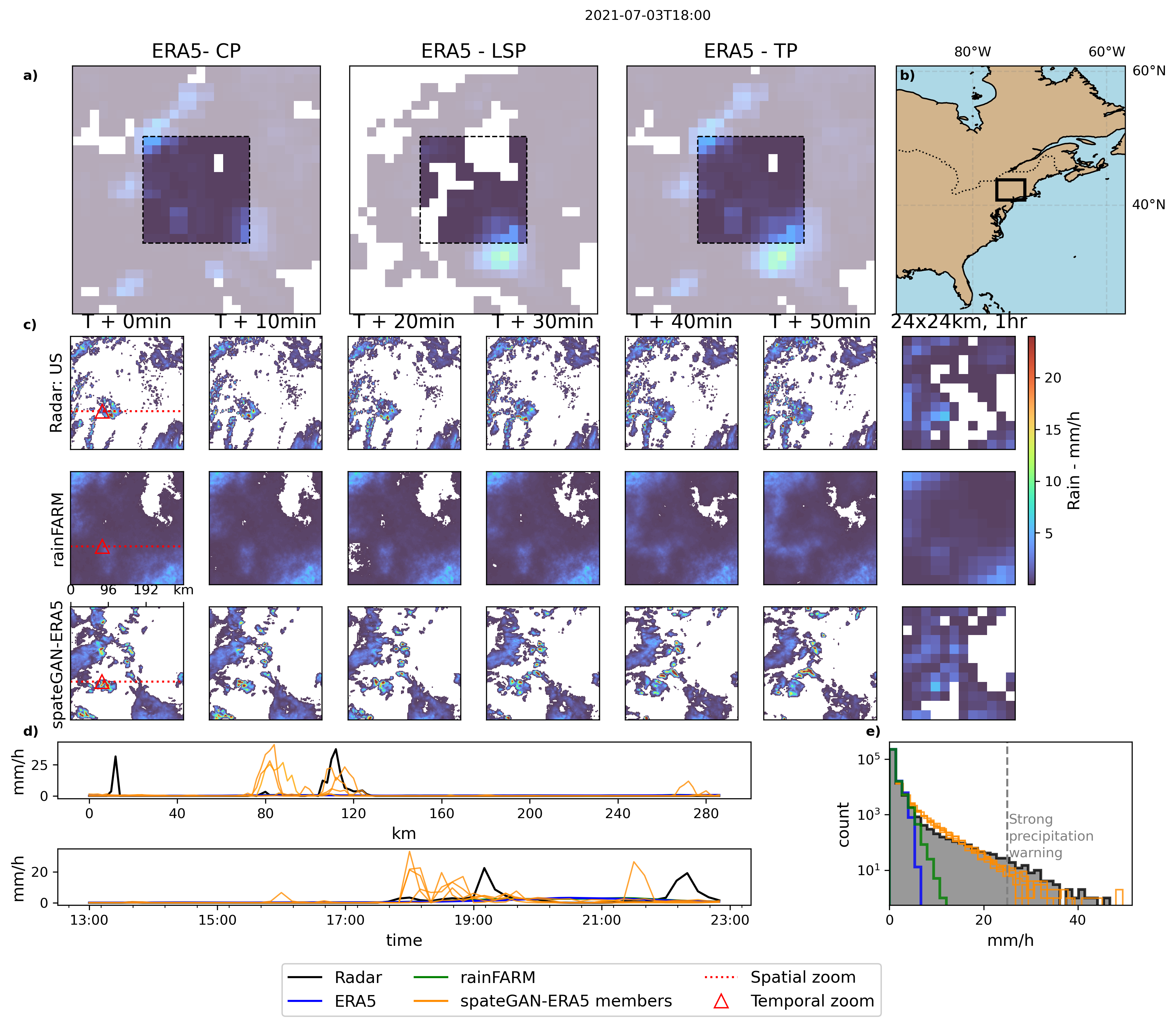}
\caption{Case study of performance on a challenging precipitation event starting on 03.07.21 in the US with observed convective cells. a) Model input patches consisting of larger ERA5 data, i.e. the convective and large-scale precipitation contribution to the total precipitation sum. b) Location of the radar observation. c) Observations, spateGAN-ERA5 predictions and rainFARM downscaling for the target domain in 10-minute increments from t to t+50 min. and as a coarsened version approximating ERA5 resolution. d) 1-D cutouts showing spateGAN-ERA5 ensemble members for a specific pixel along the temporal dimension (top panel) and a horizontal cross-section for one time-step (bottom panel). e) Distribution for temporally aggregated data with 2\,km and 1 hour resolution including maps shown in c) and the previous and following 6 hours. A severe precipitation warning threshold of the German Weather Service is set at 25 mm/h. }\label{fig2}
\end{figure}
\noindent
We select a variety of meteorologically interesting events (Fig. \ref{fig2} and supplementary Figs. \ref{sup_fig9}, \ref{sup_fig10}, and \ref{sup_fig11}) to showcase the spatio-temporal downscaling performance of spateGAN-ERA5 and how this overcomes inherent limitations of ERA5 precipitation data.\\
\\
Here, we focus on the event in Fig. \ref{fig2} showing convective cells in the United States as observed by the MRMS dataset, which are at a scale known not to be resolvable by ERA5 \cite{kendon_convection-permitting_2017}. Even when compared to coarsened radar observations at ERA5 resolution, ERA5 shows a too-low variance (see Fig. \ref{fig2}e) with an underestimation of extreme values (see further evaluation section \ref{sec6_63}).
Being able to reconstruct such small-scale rainfall cells is of particular interest to improve ERA5 precipitation estimates in regions and seasons with a high amount of convective precipitation, such as the tropics and extratropics \cite{lavers_evaluation_2022}.\\
\\
SpateGAN-ERA5 is able to reconstruct convective rainfall fields with small-scale structures and plausible rain rates, including heavy local rainfall. The rain cells show temporal continuity, hardly allowing for a qualitative differentiation between observed and predicted rainfields (see video V1 in ancillary files). Predicted rainfall may occur at a misplaced spatial or temporal position, but with a magnitude similar to the associated radar observation (see Fig. \ref{fig2}d). This misplacement is not solely due to the under-determined nature of the downscaling problem but also reflects differences between ERA5 and radar data on a coarser scale. The probabilistic nature of spateGAN-ERA5 accounts for such uncertainties, but is also constrained by the contextual information provided by ERA5. For example, the predicted ensemble shows greater variability in intensity than in spatial or temporal localization.\\
\\
RainFARM fails to reconstruct small-scale convective cells, overestimates the spatial extent of rainfall, and underestimates extremes. By design, rainFARM mostly coincides with ERA5 at the coarse resolution, limiting spatio-temporal disaggregation. This leads to only slightly more granular rainfall fields than using simple interpolation techniques. Note that spateGAN-ERA5 is the only presented method with a distribution similar to the observations and the only method that predicts larger rainfall intensities in the severe weather warning range (Fig. \ref{fig2}e), demonstrating its potential for risk analysis.

\section{Skilful representation of extreme values}\label{sec4}

\begin{figure}[ht]
\centering
\includegraphics[width=1\textwidth]{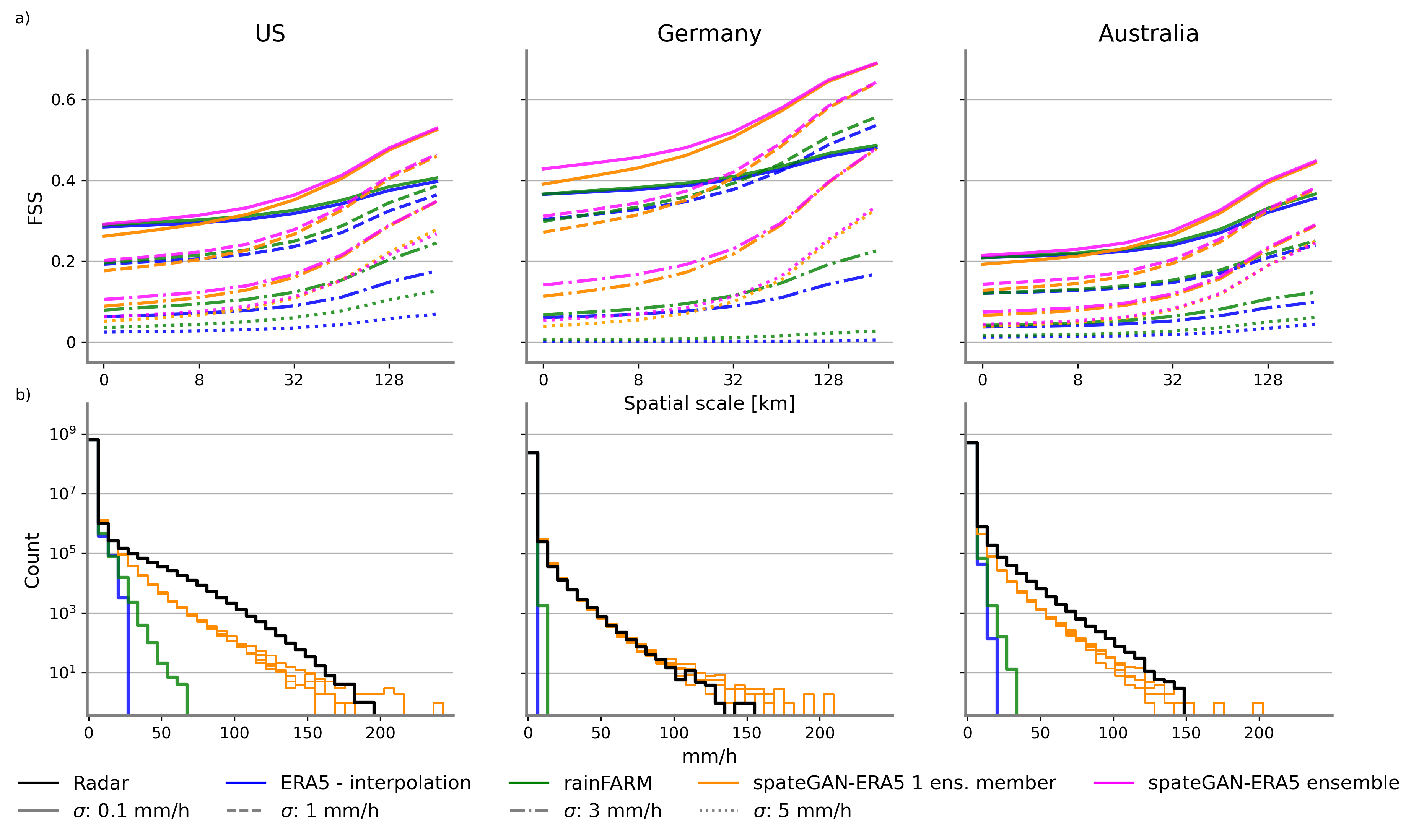}
\caption{Investigation of the downscaling distribution reconstruction skill for the evaluation datasets in Germany, US and Australia in 2021. a) Fractions skill score (FSS) for thresholds 0.1, 1, 3, and 5 mm/h and a temporal scale of 1 hr. We report the slightly improved probabilistic ensemble FSS for rainFARM. b) Distribution comparison showing multiple spateGAN-ERA5 ensemble members. }\label{fig3}
\end{figure}

To get a more complete picture of the extreme value statistics of spateGAN-ERA5 we analyse data from different climatic regions in the US, Germany and Australia (see section \ref{sec6_2}).\\
\\
The fractions skill score (FSS) (Fig. \ref{fig3}a) shows that only for the smallest rain rate threshold and up to a spatial scale of 16\,km the interpolated ERA5 and rainFARM rainfields have a higher location accuracy than a single ensemble member of spateGAN-ERA5. Considering an increased spatial scale or ensemble of predictions, the generative model consistently outperforms the other methods across all rain rate thresholds. For intense rainfall larger than 5 mm/h, spateGAN-ERA5 is the only model with acceptable skill. The relative improvement in terms of $\Delta$mFSS when considering spateGAN-ERA5 as a downscaling technique instead of an interpolated ERA5 is highest for Australia, the dataset where interpolation has the lowest absolute mFSS. It is followed by Germany, the training region, and the US (see Tab. \ref{tab1}). This indicates a strong ability of the model to generalise well outside its training domain.\\
\\
The distributions shown in Fig. \ref{fig3}b) further support spateGAN-ERA5's capability in predicting plausible extreme values. Predictions generally follow the reference’s log-normal distribution for Australia and Germany, which is physically reasonable \cite{bell_space-time_1987, crane_space-time_1990}. Different characteristics in the US are physically more implausible and thus likely due to systematic errors in the non-gauge-adjusted MRMS radar data. Overall, spateGAN-ERA5 underestimates the frequency of strong precipitation for Australia and the US and overestimates for Germany. Since spateGAN-ERA5 follows the average precipitation amount of ERA5 (see section \ref{sec5_1}), this is in agreement with the biases of the individual evaluation datasets as shown in Tab \ref{tab1}.\\
\\
In terms of a pixel-wise deterministic skill (MAE and RMSE), ERA5 interpolation and rainFARM show the best results (Tab \ref{tab1}). However, this is mainly due to their tendency to produce smoother rainfields with dampened extreme values, avoiding a double penalty for misplaced small-scale events. Since we aim for sharp probabilistic estimates, we use these scores with caution. The superior CRPS shows that spateGAN-ERA5 predictions have the highest ensemble skill. The ensemble quality, important for a correct representation of extremes, is analysed by rank histograms (Fig-sup \ref{sup_fig6}). It shows a well-calibrated ensemble with a slight under-dispersive tendency for spateGAN-ERA5 and an unfavourable heavy under-dispersive tendency for rainFARM.

\section{Spatial plausibility of highly resolved rainfall fields}\label{sec5}

\begin{figure}[ht]
\centering
\includegraphics[width=1\textwidth]{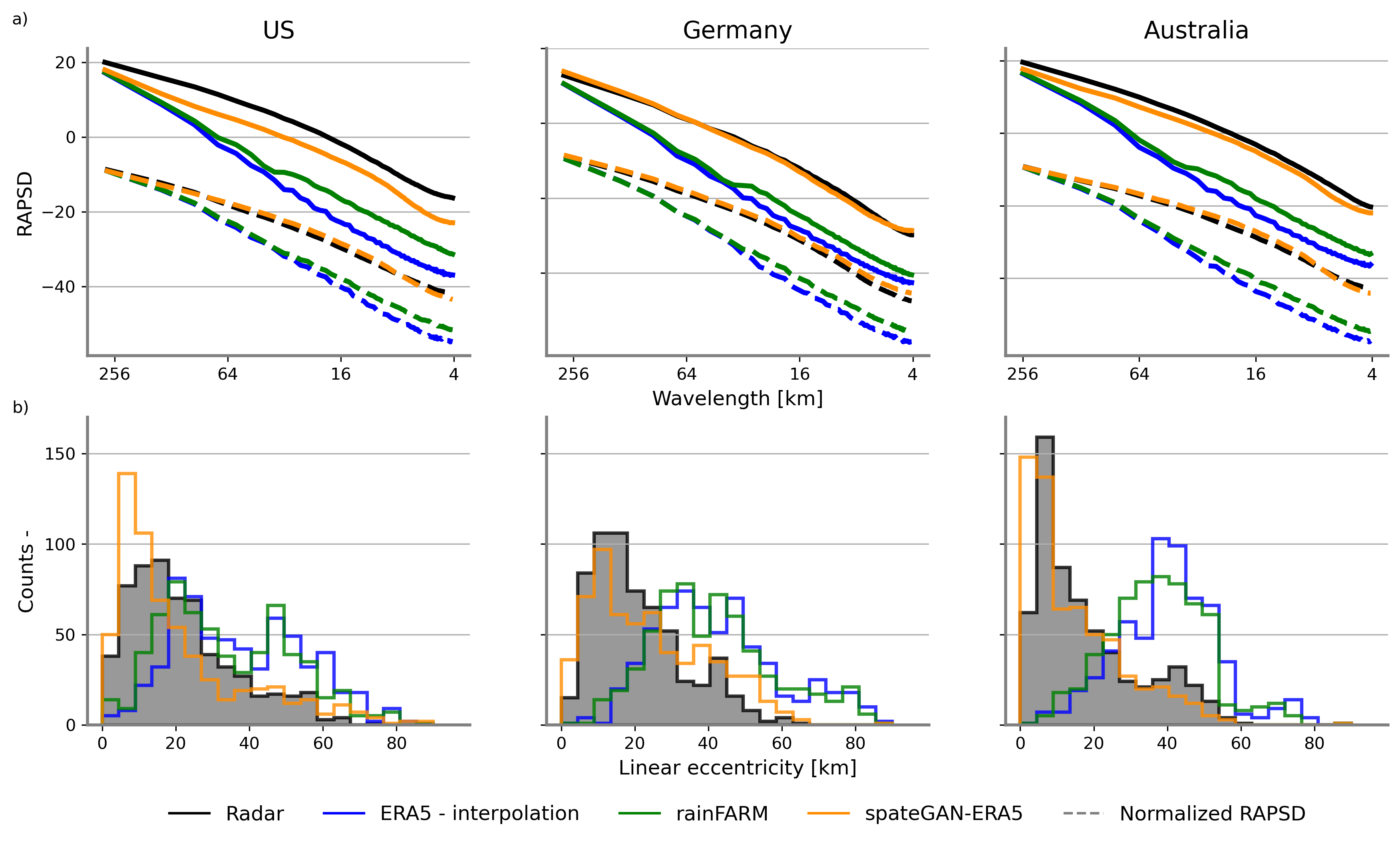}
\caption{Spatial characteristic scores for a subset of the evaluation datasets in Germany, US and Australia in 2021 (see description in section \ref{sec5_7} ). a) Mean radially averaged power spectral density (RAPSD) and mean normalized radially averaged power spectral density (dashed line). b) Distribution of linear eccentricity of the 2D autocorrelation representation (0.5 Pearson correlation coefficient ellipse).}\label{fig4}
\end{figure}

Spatial and temporal patterns of rainfall are the tangible result of the physical processes that drive precipitation formation and evolution in the atmosphere \cite{zick_shape_2016, schertzer_physical_1987}. Accurately reconstructing these patterns presents a considerable challenge, especially when using models like spateGAN-ERA5, which lack a-priori knowledge of the underlying atmospheric physics. We consider weather radar observations as a sufficient reference to allow for the statistical analysis of such spatio-temporal patterns. The qualitative assessment of section \ref{sec2} suggests that spateGAN-ERA5 predictions are hardly distinguishable from real radar observations while ERA5 interpolation produces blurry rainfields. Visually, rainFARM only slightly improves over the interpolation. To quantify this observation, we chose radial averaged power spectral density (RAPSD) (section \ref{sec6_y}). As a measure of anisotropy, a key aspect of specific spatial patterns often caused by horizontal advection \cite{schertzer_physical_1987}, we define the linear eccentricity in terms of spatial autocorrelation in section \ref{sec6_y}.\\
\\
This analysis uses a subset of each evaluation dataset, described in \ref{sec5_7}, focusing on cases with greater consistency between ERA5 and radar observations.
For the RAPSD, shown in Fig. \ref{fig4}, spateGAN-ERA5 largely replicates the power spectrum of the radar observations in Germany, with slight deviations at the smallest wavelengths close to the target resolution. In the US and Australia, an underestimation of all wavelengths is apparent. These discrepancies in RAPSD can be traced back to the mean field biases of ERA5, which are stronger for more extreme events \cite{lavers_evaluation_2022}, and by design not corrected by spateGAN-ERA5.
When focusing solely on spatial characteristics and disregarding a multiplicative bias, the normalized RAPSD shows an almost perfect alignment between predictions and observations for all datasets.\\
\\
ERA5 interpolation produces overly smoothed rainfields, resulting in a considerably lower RAPSD and normalized RAPSD for shorter wavelengths. RainFARM slightly improves the power spectrum, increasing the amplitude for wavelengths between the ERA5 resolution of 24\,km up to the final 2\,km resolution. However, the method introduces a physically unrealistic jump in the power spectrum at 24\,km \cite{crane_space-time_1990, pegram_high_2001}. 
The temporal power spectrum density shows a similar behaviour of all methods for the temporal dimension (Fig. \ref{sup_fig7}).\\
\\
Linear eccentricity is analyzed in Fig. \ref{fig4}b) and illustrated for a single field in Fig. \ref{sup_fig8}. The spateGAN-ERA5 distribution of the score is close to the observations while rainFARM stays similar to the ERA5 interpolation, providing rainfields that are highly autocorrelated for a large spatial lag. SpateGAN-ERA5 produces small-scale features that resemble the radar observations in terms of size, orientation, and eccentricity (see Fig. \ref{sup_fig_ecc}). 
\\\\
In summary, the RAPSD, temporal PSD and linear eccentricity between ERA5 and radar data support the observations from the case studies and show high structural similarities between the spateGAN-ERA5 output and the reference datasets (Fig. \ref{fig4}). A mere extrapolation of the spatial power spectrum of ERA5 like performed by rainFARM proved to be insufficient for the given problem.

\section{Conclusion}\label{sec_conc}
SpateGAN-ERA5 is unique in its ability to disaggregate and reconstruct the statistical properties of rainfields across temporal and spatial scales, with plausible extreme values, that are completely missing within the initial input data. It is a probabilistic, yet computationally very efficient, downscaling method, that accounts for the downscaling-related uncertainties when disaggregating rainfall in space and time. Event-based analyses show that, especially in convective settings, spateGAN-ERA5 skilfully reconstructs extremes and thus presents a suitable tool providing data for impact analysis and flood risk assessment.\\
\\
In this study, we demonstrate a spatio-temporal downscaling that enhances ERA5 precipitation estimates to an unprecedented spatial resolution of 2\,km and a temporal resolution of 10 minutes. However, the methodology behind spateGAN-ERA5 is generic and can be applied to other precipitation datasets and resolutions. On a more abstract level, we demonstrate that spateGAN-ERA5 is capable of learning the target distribution for loosely paired and biased input and target data. For spatio-temporal downscaling of precipitation, a variable that is notoriously difficult to model, this has not been achieved before with such a high quality.\\
\\
We validated the model with data from diverse climate zones across the US, Germany and Australia. Therefore, we are confident that the model´s generalization capabilities extend beyond these regions and can deliver high-resolution precipitation products with improved rainfall distributions on a global scale.
The abundance of high-resolution precipitation maps that can be generated several orders of magnitude faster than using dynamical downscaling addresses a critical need within the hydrological and meteorological research community, with potential applications in hydrological modelling, AI-driven weather forecasting, and beyond.

\newpage

\section{Online Methods}\label{sec6}

SpateGAN-ERA5 performs spatio-temporal downscaling of ERA5 precipitation estimates increasing the resolution from 24\,km and 1 hr to 2\,km and 10 min. The model receives input patches of the ERA5 variables convective and large-scale precipitation of size 16 hr x 672\,km x 672\,km and performs the downscaling for a  centred domain of 8 hr x 336\,km x 336\,km. We trained the model in Germany where a consistent high-resolution and high-quality reference dataset is available through the gauge-adjusted and climatology corrected radar product RADKLIM-YW provided by the German Meteorological Service and where a high agreement between ERA5 precipitation and observation data can be shown (see Fig. \ref{fig3}) \cite{lavers_evaluation_2022}. Global downscaling is achieved by downscaling and stitching of overlapping patches.

\subsection{Model description}\label{sec5_1}
\begin{figure}[ht]
\centering
\includegraphics[width=1\textwidth]{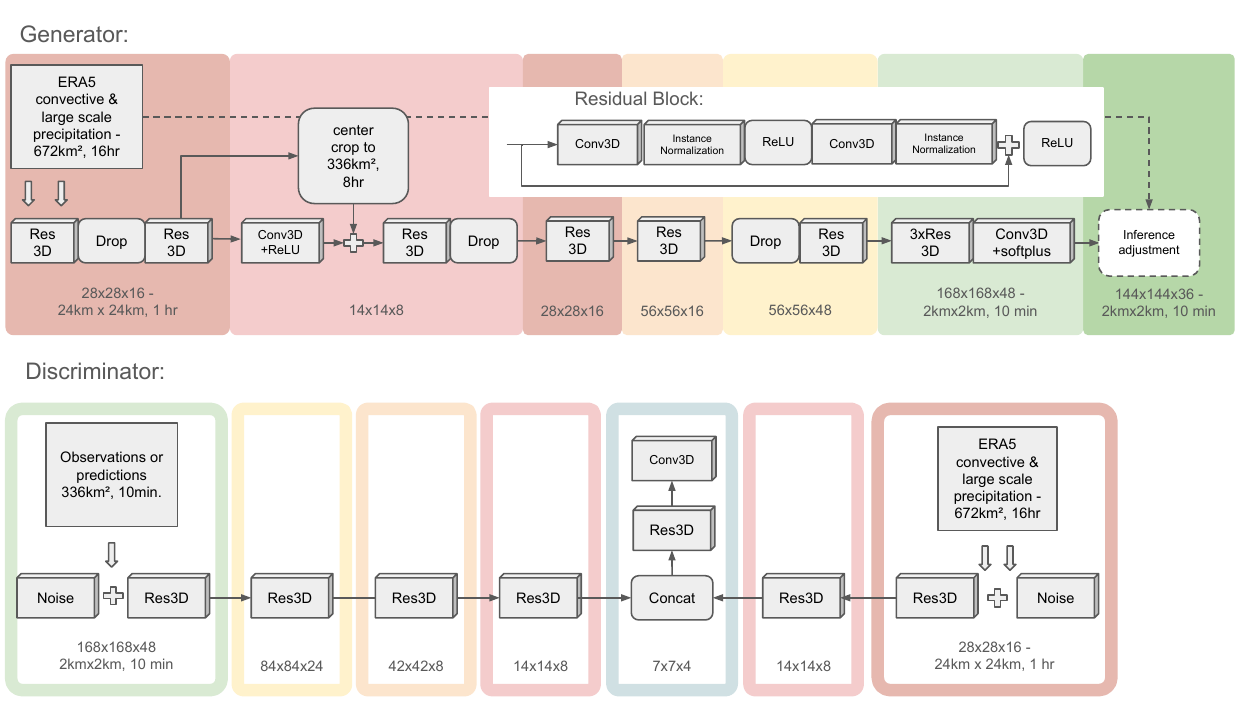}
\caption{Schematic overview of the spateGAN-ERA5 generator and discriminator model architecture using Residual Blocks (Res3D). During inference boundary regions are removed and an ERA5 mean field bias adjustment is applied.}\label{fig5}
\end{figure}
\noindent
We build on the successful precipitation video super-resolution approach, spateGAN \cite{glawion_spategan_2023}, which makes use of 3D convolution. The main ERA5 downscaling model consists of four consecutive components that make use of 3D-convolutional residual blocks (Res3D)  to better capture spatio-temporal dependencies and to improve convergence during training.\\
\\
First, the ERA5 convective and large-scale precipitation input data is processed on its initial resolution, afterwards, it passes a UNET-like downsampling and skip connection with an added cropping operation. This allows the model to process data at multiple resolutions, consider global and local features and focus on the target domain at an early model stage. Afterwards, the spatial and temporal resolution of the input data is successively increased and the structures of the rainfields are refined by 4 upsampling blocks, including bilinear and linear interpolation and Res3D blocks. Three subsequent Res3D blocks adjust fine-scale structures and limit the prediction range to positive values using a final Softplus activation function.\\
\\
Temporally constant dropout (p=0.2) at three different generator depths introduces scale and rain event-dependent perturbation at low, mid and high frequencies and enables spatio-temporally continuous probabilistic downscaling. The perturbation in combination with the ensemble loss supports the model in reconstructing the missing tail of the ERA5 precipitation distribution.\\
\\
During inference mode, i.e. for model selection, evaluation and global prediction, we apply three additional operations. First, we freeze one dropout seed for each produced ensemble member, which improves the spatio-temporal consistency of the rainfields compared to a random perturbation in space and time. Second, we cut the outermost edges (24\,km and 1 hr) to remove boundary effects. For global predictions, this routine differs slightly, as described in \ref{sec5_8}. Third, we apply a patch-wise mean field bias correction to the predictions \cite{hess_physically_2022} to ensure that the provided patch-wise average ERA5 precipitation amount is preserved and that the model can be applied in regions where the ERA5 bias strongly deviates from the training distribution. \\
\\
The overall design of the generator is memory efficient and can be run in inference on smaller GPUs (10 Gb per sample). This allows the application of our model by a broad research community, not only those with access to the latest-generation GPUs with large memory. \\
\\
The discriminator (see Fig. \ref{fig5}) is trained simultaneously with the generator. Its inputs are the temporal sequences of high-resolution prediction or observation, as well as the coarse-resolution context provided to the generator. Its training objective is to decide if the high-resolution field is real, or artificially generated. The loss function is binary cross-entropy. Data is processed by a series of Res3D blocks to process spatio-temporal patterns followed by strided convolutions. The processed coarse and high-resolution inputs are concatenated at a late stage to encourage a comparison based on latent features extracted on multiple resolutions. The discriminator model is thereby used as a powerful dynamical loss function for the generator, that learns to discriminate structure- and distribution-related rainfall characteristics.

\subsection{Training and model selection}\label{sec5_2}

The model is trained for $2\times10^{5}$ adversarial training steps. The learning rate is $1\times 10^{-4}$ for the generator and $2\times10^{-4}$ for the discriminator and uses AdamW optimizer \cite{loshchilov_decoupled_2019} with $\beta_1=0.0$ and $\beta_2=0.999$ (Discriminator: $\beta_1=0.0$ and $\beta_2=0.5$). We employ data-parallel training on 3 Nvidia A100 GPUs for 4 days. The batch size is set to 9 per training step. In inference mode, downscaling 1 patch takes 0.04 s on one A100 GPU.\\
\\
We save all model weights after every 250 training steps and identify the best generator training state by downscaling and evaluating the independent model selection dataset \ref{sec5_6}. We select the final model by calculating the average of the ensemble FSS (meFSS) of the thresholds 0.1, 1, 3, 5 and 8 mm/h, spatial scales 1, 4, 8, 16, 32, 64 and 128\,km and temporal scale of 1 hour. This considers the ensemble quality and location accuracy for different categories of rainfall intensities, independent of the heavily skewed distribution of rainfall.

\subsection{Evaluation}\label{sec5_4}

For evaluation, we calculate a set of quantitative scores since no single metric is capable of capturing the complexity of highly resolved rainfields. For this, we investigate the FSS, mFSS, meFSS, RAPSD, CRPS, RMSE, MAE, rank histograms and spatial autocorrelation (see section \ref{sec6_y}) for data on a regular km grid. During evaluation, spateGAN-ERA5 downscales patches that overlap in the temporal dimension to generate a continuous sequence of temporally consistent rainfields is, by keeping the central 2 hours of each patch. 
For the case study videos, a linear blending approach is applied to 1 hour overlapping periods, with weights decaying from 1 to 0, effectively smoothing out minor remaining temporal discontinuities in the predictions. 
\\
\\
In total, the probabilistic model performance is evaluated using 100 ensemble members for calculating rank histograms and CRPS shown in Fig. \ref{sup_fig6} and Tab \ref{tab1}. For the ensemble FSS and meFSS, we calculate only 6 members since the score converges at a small ensemble size. For the presented evaluation, rain rates smaller than 0.01 of all compared datasets are set to zero.

\subsection{Datasets}\label{sec6_2}
The model input and, therefore, the only dataset required for applying spateGAN-ERA5 are the convective and large-scale variables from the ERA5 reanalysis. The model is trained using gauge-adjusted and climatology-corrected radar data in Germany. We use two additional radar datasets for evaluation from the United States and Australia to test the model's ability for generalisation outside of its training distribution. Even if it seems obvious at first to include data from the US and Australia for model training, we have deliberately refrained from doing so. Pure radar observations can be highly error-prone and do not match the quality of a sophisticated, gauge-adjusted and climatologically corrected product such as RADKLIM-YW. Due to the lack of high-resolution data availability, we use radar observations to get an indication of spateGAN-ERA5's generalisation capabilities.

\subsubsection{ERA5 dataset}\label{sec6_21}

The ERA5 reanalysis provides global, hourly model data spanning the past 70 years \cite{hersbach_era5_2020, bell_era5_2021}. It integrates observational data with numerical model predictions through advanced data assimilation techniques, resulting in a high-quality benchmark dataset. 
For precipitation, the  ERA5 4D-var system assimilates hourly NCEP stage IV gauge-adjusted weather radar precipitation information over the US \cite{lopez_direct_2011, lin_ncep_2005}.
In this study, we used the years 2009 to 2021 where ERA5 aligns with the available radar data.
We utilise the variables convective and large-scale precipitation of hourly ERA5 data as input for spatio-temporal downscaling. Including additional variables as input, such as wind components, temperature, pressure level etc., did not enhance overall performance in the presented setup. \\
\\y lack valuable scale-related information \cite{zandler_evaluation_2019, chen_spatial_2021, gomis-cebolla_evaluation_2023}, excludes oceans and coastal areas and has a higher release latency \cite{xu_era5_2022}.\\
\\
Despite the known limitations of ERA5 precipitation estimates, which include spatially heterogeneous quality, biases \cite{hersbach_era5_2020, lavers_evaluation_2022}, a tendency to smooth out local extremes due to the coarse resolution of 0.25° and 1 hour \cite{bandhauer_evaluation_2022}, and limitations in modelling convective events, \cite{beck_mswep_2019, kendon_convection-permitting_2017} the product is most commonly used in environmental research.

\subsubsection{RADKLIM-YW Germany}\label{sec6_22}

For training, model selection and part of the validation of spateGAN-ERA5 we use the gauge-adjusted and climatology-corrected weather radar product RADKLIM-YW provided by the German Meteorological Service (DWD) as target data \cite{winterrath_erstellung_2017, winterrath_radar_2018}.\\
\\
This product is a composite of precipitation information from a network of 16 C-band weather radars.
It is adjusted by approx. 1000 rain gauges homogeneously distributed in Germany with a density of one gauge per 330\,km$^2$. In addition to the RADOLAN gauge adjustment, effects like range-dependent underestimation and beam blockage are covered by an additional climatological correction.\\
\\
The grid extent is 900\,km × 1,100\,km in polar stereographic projection, covering almost entire Germany and its surrounding border regions, with a resolution of 1\,km × 1\,km and a temporal resolution of 5 min. Each grid cell represents a 5 min. rainfall sum with a quantization of 0.01 mm. Regions not covered by the 150\,km measurement radii of the radars or missing measured values are marked with “NaNs.” For our investigation, we used data on the provided\,km grid, coarsened to 2\,km and 10 min. resolution. We use the years 2009-2020 for model training, the first half of the year 2021 for model selection and the second half for evaluation, preventing data leakage and testing for generalisation abilities. For evaluation, we select two fixed locations of the size 336\,km x 336\,km, highlighted in Fig. \ref{fig1}   covering almost the entire country. 
\\
\subsubsection{Multi-Radar Multi-Sensor System (MRMS) United States}\label{sec6_23}

For validation purposes, we use the radar composite from the Multi-Radar Multi-Sensor (MRMS) system comprising 146 WSR-88D radars covering the US and 30 Canadian radars \cite{zhang_multi-radar_2016, smith_multi-radar_2016}. Climatic conditions in the United States have a high variability ranging from continental, subtropic and Mediterranean to tropical.\\
\\
The MRMS dataset we use is not gauge adjusted. Alternative gauge-adjusted QPE products are not available at a sub-hourly resolution and, therefore, not suitable for most parts of our analysis.\\
\\
MRMS covers the region from 20° to 55° latitude North and 130° to 60° longitude West with a resolution of 0.01° in both latitude and longitude direction. The temporal resolution is 2 minutes.
We select 6 regions exhibiting a high radar quality and covering different climatic regions of the country (see Fig. \ref{fig1}, yellow boxes).  For evaluation, we regrid the radar observations of each 6 locations to their associated regular km UTM projection and downsample it to 2\,km and 10 min. resolution. Each location has a domain size of 336\,km x 336\,km.

\subsubsection{Australian Radar Network}\label{sec6_23x}

We additionally use quantitative precipitation estimates from the Australian operational radar network \cite{seed_aura_2022}. \\
\\
We select data from 6 different C-band weather radars which cover sub-tropical regions in the south and tropical regions in the north of the country (see Fig. \ref{fig1}). The individual locations have a radar coverage of 150\,km and are selected by considering less beam blockage, data availability and homogeneous distribution. 3 of these radar sites operate Doppler radars.
The QPE is gauge-adjusted but strongly depends on the availability of the heterogeneously distributed rain gauge observations  \cite{chumchean_correcting_2006}. An increased bias between ERA5 and the Australian radar was visible and the radar quality may be a larger factor than in Germany (see Tab. \ref{tab1}).
The product has a spatial grid resolution of 0.5\,km x 0.5\,km using an Albers Conical Equal Area projection and a temporal interval between 5, 6 and 10 min. For evaluation, we downsample the observations to 2\,km and 10 min. resolution. Due to the smaller radar coverage, each location has a domain size of 280\,km x 280\,km.

\subsection{Data preparation}\label{sec5_5}

Observation data and ERA5 precipitation estimates are adjusted to be used for model training, selection, evaluation and global inference as described below.

\subsubsection{Training and model selection dataset}\label{sec5_6}

For training, we draw random target samples from RADKLIM-YW, each with 48 continuous radar observation time steps and a size of 168 x 168 pixels, i.e. 8 hours and 336\,km x 336\,km. The associated model input is received by first interpolating ERA5 data to the target grid and afterwards downsampling the extracted patches to 24\,km and 1 hr to approximate the initial resolution.\\
\\
We apply a subsampling routine, selecting only samples with a sufficient amount of wet pixels and total precipitation in both input and target to avoid learning from data that contains little to no rain and fewer wet pixels (see section. \ref{sec6_4}). The resulting number of observation samples contained in the trainings data is about 20000 (850 GB).\\
\\
For model selection we randomly draw additional samples and apply them to the subsampling routine. We select 1000 samples from the temporally independent time period (January to June 2021). 
We adjust the average rainfall of the targets of this dataset, using a scalar multiplication, so that it matches the average rainfall of the corresponding ERA5 data. This supports the identification of a model state that tends to modify the average precipitation of the ERA5 input samples less drastically and allows the model to be applied outside the training region.

\subsubsection{Evaluation dataset}\label{sec5_7}

SpateGAN-ERA5 is evaluated using a temporally and spatially independent dataset. The evaluation period contains every first week of the months of July to December 2021. The data is sampled using fixed patch locations in the US, Germany and Australia, highlighted in Fig. \ref{fig1}. For the associated ERA5 samples, the data are projected to the observations grid and afterwards interpolated to 24\,km resolution. The domain size is 672\,km and includes the previous and following 8 hours of the evaluation observation time period. \\
\\
To analyze the spatial characteristics of the predicted rainfields, i.e. radially averaged power spectral density and anisotropy, we select a subset of each evaluation dataset that exhibits greater consistency between ERA5 and radar observations. This subset includes cases where interpolated ERA5 achieves a mFSS score exceeding 0.2. \\
\\
In addition to the high-resolution observations and predictions, we evaluate the performance of the individual downscaling methods and datasets on a coarser resolution, approximately the one of ERA5 (see section \ref{sec6_63}). Therefore we average the observations and predictions of the evaluation datasets to a spatial resolution of 24\,km using 2D average pooling and aggregate the temporal dimension to 1 hour resolution.

\subsubsection{Generation of global fields}\label{sec5_8}

We define a processing pipeline for producing seamless global high-resolution precipitation maps, from a deep learning model that operates on patchwise downscaling.\\
\\
First, ERA5 data on its original lat-lon grid is segmented into patches. Each patch covers a regular spatial extent of 672\,km x 672\,km. We calculate the necessary ERA5 lat-lon coordinates to maintain these patches with the required spatial extent by using the Haversine formula. To simplify the process, the latitude centre coordinate of each patch is used to determine the longitudinal extent. Resulting spatial distortions in the longitude directions can be neglected due to the small patch sizes. In comparison to the evaluation and training datasets, where ERA5 is regridded onto a regular kilometre grid using the radar observation projection or UTM projection, this is a more efficient method for global high-resolution mapping. The patches are designed to overlap, such that the target prediction domain of 336\,km x 336\,km overlaps by approximately 10\% in both latitude and longitude directions.\\
\\
The generated patches are then interpolated onto a regular grid with dimensions of 672\,km x 672\,km using nearest neighbour interpolation. This data has an approximate resolution of 24\,km and enters the spateGAN-ERA5 model as input data. Downscaling of patches on km-grid ensures that the model receives data that does not exhibit any latitude-dependent spatial distortion of physical properties. After downscaling, spateGAN-ERA5 applies a mean field bias adjustment. Due to extensive areas of uncertain, low-intensity rainfall in the ERA5 dataset - particularly over ocean regions - all ERA5 rain rates below 0.1 mm/h are set to zero for this adjustment. The resulting downscaled high-resolution patches are seamlessly interpolated onto a global latitude-longitude grid with a resolution of 0.018 degrees, which corresponds to approximately 2\,km at the equator.\\
\\
To combine the individual overlapping patches a linear weighting (decaying from 1 to 0 while approaching the border of the patch) is applied in the overlapping regions. This blending process ensures smoother transitions between patches, aiming for continuous large-scale rainfall field circulation (see ancillary
files Fig. A1). 

\subsection{Reference methods}\label{sec5_9}

RainFARM is a statistical downscaling approach implemented in the PySTEPS package \cite{pulkkinen_pysteps_2019}. It produces small-scale variability by a stochastic process that estimates and extends the spectral slope from each coarse input patch with an estimated scaling factor while preserving key statistical properties. Most importantly, rainFARM produces an isotropic spatial distribution and preserves the rainfall amount when aggregated to the initial resolution. We apply spatial downscaling of ERA5 total precipitation using the advanced spectral rainFARM algorithm \cite{donofrio_stochastic_2014}, followed by temporal interpolation. A probabilistic downscaling is achieved using a different fixed random seed for the stochastic component of the method.\\
\\
For this particular problem, the performance was better than applying the combined spatio-temporal downscaling operation described in \cite{rebora_rainfarm_2006}. Downscaling and aggregating the individual ERA5 variables convective and large-scale precipitation separately lead to negligible differences. Unlike spateGAN-ERA5, rainFARM downscales patches of the whole ERA5 input domain of 672\,km x 672\,km and 16 hours and is afterwards cropped to match the domain of the radar observations from the evaluation datasets.\\
\\
Trilinear interpolation of ERA5 total precipitation, in both space dimensions and the time dimension, serves as a simple baseline where the ERA5 rainfall information can be compared to the high-resolution radar observation without an artificial generation of small-scale features. We interpolate the projected ERA5 data on the coarse km grid described in \ref{sec5_7}.
\\
\\
\\




\bmhead{Acknowledgements}
This work was supported by funding from the Federal Ministry of Education and Research (BMBF) and the Helmholtz Research Field Earth \& Environment within the Innovation Pool Project SCENIC and the HFMI project HClimRep. Further support has been granted through the DFG research unit RealPEP (Grant Number: CH-1785/1-2).
Furthermore, we acknowledge the support of the Deutsches Klimarechenzentrum (DKRZ) by providing high-performance cluster resources, granted by its Scientific Steering Committee (WLA) under project ID 1343.



\section*{Declarations}



\textbf{Data availability} \\
The results and model of this study are produced by publicly available datasets ERA5 \cite{hersbach_era5_2020}, RADKLIM-YW \cite{winterrath_radar_2018}, MRMS \cite{smith_multi-radar_2016, zhang_multi-radar_2016}  and Australian operational radar network \cite{seed_aura_2022}. The ERA5 dataset can be downloaded from https://cds.climate.copernicus.eu/.  The Australian observations can be accessed from https://thredds.nci.org.au/thredds/catalog/rq0/rainfields3/catalog.html. 
\\
\\
\textbf{ Code availability} 
The study was conducted using several open source frameworks, including pytorch \cite{paszke_pytorch_2019} (https://pytorch.org/) and pySTEPS (https://github.com/pySTEPS/pysteps). Maps were produced using cartopy (scitools.org.uk/cartopy). 
The spateGAN-ERA5 model, implemented and optimized in a Python framework, will be made available until publication.
\\
\\
\noindent

\bibliography{references}

\FloatBarrier
\newpage

\begin{appendices}
\section{
Supplementary information}
\label{sec7}


\backmatter


\subsection{Further methodical details}\label{sec6_3}

Here, we provide further methodical details, including information about the training data sampling scheme, the spateGAN-ERA5 model architecture, its training objective and the evaluation metrics used in this study. 

\subsubsection{Training data sampling scheme}\label{sec6_4}

Most of the time little to no rain falls in the training region of Germany. Therefore, we apply a training data sampling scheme to extract samples that contain a higher rainfall amount, more wet pixels and show a higher agreement between ERA5 precipitation and observations. 
For each randomly drawn sample, the following conditions must be fulfilled by the ERA5 input $X$ and the RADKLIM-YW observation $y$:

\begin{enumerate}
    \item $X$ and $y$ do not contain missing values
    \item $\sum_{h,w,t} X > \varepsilon_1$ and $\sum_{h,w,t} y > \varepsilon_1$, where $\varepsilon_1 = \vert -450 \varepsilon + 4500 \vert$ and where $\varepsilon$ is drawn from $Lognormal(0,1)$.
    \item The 66th quantiles of the pixel values in $X$ and $y$ exceed $\varepsilon_2$, where $\varepsilon_2 = \vert -50 \varepsilon' + 500 \vert$ and where $\varepsilon'$ is drawn from $Lognormal(0,1)$.
\end{enumerate}

\begin{figure}[ht]
\centering
\includegraphics[width=0.5\textwidth]{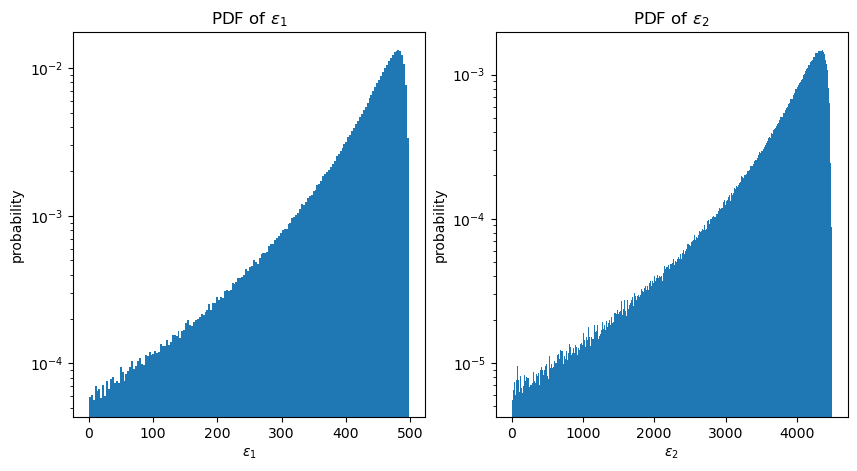}
\caption{Probability density of $\varepsilon_1$ and $\varepsilon_2$.}\label{sup_fig_d}
\end{figure}

The distribution of the thresholds $\varepsilon_1$ and $\varepsilon_2$ is shown in Fig. \ref{sup_fig_d} and roughly reflects the inverse probability of drawing samples that match the given thresholds.

\subsubsection{Model architecture}\label{sec6_5}

SpateGAN-ERA5 consists of a generator, trained in an adversarial manner with a discriminator model. A schematic overview of the model's architecture is shown in the main article in Fig. \ref{fig5}.
\\
The generator input is the low-resolution condition stack of the two ERA5 variables convective and large-scale precipitation with a spatial extent of 672\,km (28x28 pixels) and a time sequence of 16 hours. 
The task of the generator is to learn a mapping of the given context to construct rainfields how they would have been observed by a rain gauge adjusted radar observation product on a refined grid size of 2\,km and temporal resolution of 10 min. The actual prediction and target domain show half the temporal and spatial extent as the context domain. \\
\\
The generator first extracts features of the given ERA5 context using one 3D convolutional residual layers (Res3D) Block, followed by a Dropout layer and another Res3D block. From this layer, a UNET-like skip-crop connection is applied, where one path decreases the dimension of the data using strided 3D convolution, and another path center crops the data. Both outputs have the dimension 14x14 pixels and 8 hours and are added before entering one Res3D and Dropout layer. 
For downsampling operations, the Res3D blocks include a 3D convolutional layer with a kernel size of 1 and instance normalisation within the skip connection to harmonise the dimensions. Within the generator, the data is further processed by combining Res3D blocks and bilinear interpolation for the space dimension and linear interpolation for the time dimension. After four upsampling steps, the final data resolution of 2\,km and 10 min. is reached. On this scale, three additional Res3D blocks are applied. A final 3D convolutional layer with a Softplus activation function projects to one feature dimension with positive values.
During inference mode, the dropout seed is frozen for each ensemble member prediction. Additionally, another operation removes the outermost edges (24\,km and 1 hr) of the predictions and multiplies the average predicted rainfall by a single value to match the average rainfall amount of the associated ERA5 input patch.
\\\\
The discriminator likewise makes effective use of Res3D blocks in combination with strided downsampling, to process high-resolution observations or predictions and ERA5 convective and large-scale precipitation. Both, the high-resolution and low-resolution data are treated separately and, as a first step, Gaussian noise (mean=1, std. 0.05) is added to the input data to prevent the model from learning to distinguish rainfields based on quantization characteristics. In total 4 Res3D Blocks for the high-resolution path and 2 Res3D Blocks for the low-resolution path process the data until the dimensions are harmonised and concatenated. After one last Res3D Block and a final Conv3D layer, the output is used for classification. \\
\\
The generator uses 3D reflection padding in all layers with 3D convolution and the discriminator uses zero padding. Except for the first, second and last Res3D Block of the generator, the first Res3D block of the high-resolution discriminator path and the two Res3D blocks of the low-resolution discriminator path, we apply instances of normalized convolutions. All convolutional layers in the networks use a kernel size of 3. For the generator, we use a feature dimension of 96. For the discriminator, the high-resolution features are 128, 128, 128 and 64 and the low-resolution are 64, 32. After concatenating the final Res3D Block decreases the features to 64, which are compressed to 1 within the last 3D convolutional layer.\\
\\
The specific model architecture stems from an iterative optimization process that started during our investigations for precipitation video-super-resolution in \cite{glawion_spategan_2023} and was further developed for the task of ERA5 precipitation downscaling. Thereby we also tried e.g. state-of-the-art vision transformer network layers as generator, which did not result in a performance improvement and led us to stick to the well-proven 3D Residual layers. In general, an extensive hyperparameter optimization is of course desirable. Due to the computational complexity, long training runs and limited computational resources, we could not test all possible parameter combinations and therefore can not state that spateGAN-ERA5 provides the best possible results. We rather invite the research community to build up on our work and further improve downscaling of precipitation data.

\subsubsection{Objective function}\label{sec6_x}
As an objective function, we use a well-known stepwise adversarial training strategy \cite{goodfellow_generative_2014, isola_image--image_2017}.
\\
The discriminator D receives the ERA5 context $X$ and target observations $Y$ or predictions $\hat{Y}$ of the generator and is trained to minimise the binary cross-entropy loss

\begin{equation}
\mathcal{L}_D = -\mathbb{E}_{X, Y}[\log D(X, Y)] - \mathbb{E}_{X, \hat{Y}}[\log (1 - D(X, \hat{Y}))]
\end{equation}
The generator loss includes an adversarial loss
\begin{equation}
\mathcal{L}_{\text{GAN}}(G) = -\mathbb{E}_{X}[\log D(X, G(X))]
\end{equation}
and an ensemble L1-loss defined as
\begin{equation}
\mathcal{L}_{\text{L1}}(G) =  \overline{ \left| Y - \frac{1}{3} \sum_{i=1}^{3} \hat{Y}_{i} \right|},
\end{equation}
which compares high-resolution targets to the ensemble mean prediction of 3 members \(\hat{Y}_1, \hat{Y}_2, \hat{Y}_3\). This enures that the predictions remain close to the ground truth while reducing a double penalty of small convective cells or heavy precipitation misplaced during training.
\\\\
The total generator loss is
\begin{equation}
    \mathcal{L}_G = \mathcal{L}_{\text{GAN}}(G) + \mathcal{L}_{\text{L1}}(G)
\end{equation}

\subsubsection{Evaluation metrics}\label{sec6_y}
	
We verify the performance of the downscaling methods using a set of common metrics.
\\\\
The \textbf{root mean square error (RMSE)} is a pixel-wise error computed for a single predicted ensemble member:
\begin{equation}
RMSE =  \sqrt{\overline{(Y - \hat{Y}_i)^2}}
\end{equation}
\\\\
The \textbf{continuous ranked probability score (CRPS)}\cite{gneiting_strictly_2007} measures the prediction accuracy by accounting for the ensemble spread and bias. The Cumulative Density Function (CDF) of the predicted ensemble at a specific point and time step ($\hat{F}(xt)$) is compared to the observed rainfall $y$.

\begin{equation}
    \begin{aligned}
CRPS(\hat{F}, y) = \int_{\infty }^{-\infty } (\hat{F}(xt) - 1(xt\geq y))^2 dxt \\
    \end{aligned}
\end{equation}

\begin{align}
\ & 1(xt\geq y) \mapsto \begin{cases}
0 : & xt < y\\
1 : & xt \ge y
    \end{cases}
\end{align}
We report the CRPS as the average CRPS for each dataset. For deterministic methods (ensemble size of 1),  i.e. for interpolated ERA5, this score reduces to the \textbf{mean absolute error (MAE)}.
\begin{equation}
MAE = \overline{\left|{(Y - \hat{Y}_i)}\right|}
\end{equation}
\\\\
The \textbf{fractions skill score (FSS)} \cite{roberts_assessing_2008, roberts_scale-selective_2008} is defined as

\begin{equation}
FSS = 1- \frac{\overline{(f_{\hat{Y}}-f_Y)^2}}{\overline{{f_{\hat{Y}}}^2}+\overline{f_Y^2}},
\end{equation}

where $f_{Y}$ (resp. $f_{\hat{Y}}$) is the fraction of pixels within a spatial and temporal ($s,t$) neighbourhood that exceed a certain observed (resp. predicted) rainfall intensity threshold ($\sigma$). The averaging is performed over the respective neighbourhoods of all locations and timesteps of each evaluation dataset. For the ensemble FSS, the fraction of ensemble members exceeding ($\sigma$) is considered.\\
\\
We calculate the mean FSS (mFSS) or mean ensemble FSS (meFSS) of a set of different scales (s=0, 4, 8, 16, 32, 64, 128, 256\,km, t=1hr) and thresholds ($\sigma$=0.1, 1, 3, 5 mm/h). 
The $\Delta$mFSS is the relative deviation of the meFSS of rainFARM and spateGAN-ERA5 to the mFSS of interpolated ERA5, expressed as a percentage, and illustrates the performance benefits when considering an alternative downscaling method instead of pure interpolation. For data on ERA5 resolution, the mFSS considers spatial scales of 0, 24, 96 and 192\,km and rain thresholds of 0.1, 1, 3 and 5 mm/h.
\\
\\
The \textbf{radially averaged power spectral density (RAPSD)} and \textbf{power spectral density (PSD)} \cite{harris_multiscale_2001, sinclair_empirical_2005} measures how power is distributed across spatial and temporal frequencies. The temporal PSD acts thereby as an indicator for plausible advection. 
The RAPSD is calculated for single images  using the PySTEPS \cite{pulkkinen_pysteps_2019} implementation and is averaged for each evaluation dataset. The PSD is calculated along the temporal dimension for each pixel and for each week of the evaluation datasets and is afterwards averaged for each dataset. Additionally, we report the normalized RAPSD and PSD, where the power spectrum of each image or time sequence is normalized so that it sums to one.
\\
\\
We use \textbf{rank histograms} \cite{candille_evaluation_2005, hamill_interpretation_2001} to validate the variability and reliability of an ensemble of probabilistic rainfall predictions. For each pixel and time step of the evaluation datasets, 100 ensemble predictions are considered in increasing order and the normalized rank $r$ of the actual observation value is determined. Perfect calibrated ensembles show a uniformly distributed $r$, where predictions and observations stem from the same distribution.
\\
\\
We investigate the spatial anisotropy of rainfields by calculating the autocorrelation of single images of observations and predictions for spatial lags from 0 to 60\,km in x and y direction \cite{foresti_quest_2024}. We estimate an ellipse from the 0.5 Pearson Correlation Coefficient (PCC) counterline for each individual autocorrelation field and retrieve the variables length of major axis $a$ and length of minor axis $b$  to determine the \textbf{linear eccentricity} 
\begin{equation}
ecc_l = \sqrt{a^2-b^2},
\end{equation}
\textbf{eccentricity}
\begin{equation}
ecc = \sqrt{1-  \frac{b^2}{a^2}},
\end{equation}
and \textbf{size}
\begin{equation}
size = \sqrt{a*b}.
\end{equation}

Furthermore, we compute the \textbf{orientation} of the ellipse, i.e. of the major axis, in degrees.
\\
\\
We define the \textbf{BIAS} as
\begin{equation}
    BIAS = \frac{\overline{Y}-\overline{X}}{\overline{Y}}
\end{equation}
where $\overline{X}$ is the average predicted precipitation amount of each evaluation region and $\overline{Y}$ is the average observed rainfall.

\subsection{Further evaluation}\label{sec6_6}

Here, we provide a supplementary evaluation for spateGAN-ERA5. Tab \ref{tab1} shows a set of quantitative scores. The bias of all three methods is expected to be similar due to the nature of the methods. Slight deviations are related to rainFARMs Gaussian smoothing operation or that the bias is computed over the centre 2 hours of the predictions, but the mean field bias correction for spateGAN-ERA5 is applied to the whole 6 hour patch.

\begin{table}[h]
\label{tab1}
\centering
\caption{Quantitative performance analysis of the evaluation datasets for Germany, US and Australia and for the comparison methods spateGAN-ERA5 (SG), rainFARM (RF) and linear interpolation (LI). CRPS is shown for the probabilistic models. $\Delta$mFSS compares the model skill deviation to the interpolation skill. 
}
\begin{tabular}{|l|ccc|ccc|ccc|} \hline  

\textbf{} &  \multicolumn{3}{|c|}{\textbf{RADKLIM}} &  \multicolumn{3}{|c|}{\textbf{MRMS}}  & \multicolumn{3}{|c|}{\textbf{Australia}}  \\
\text{}         & \text{SG} & \text{RF} &  \text{LI}    & \text{SG}    & \text{RF} &  \text{LI}    & \text{SG}  & \text{RF} &  \text{LI}\\ \hline  

\textbf{CRPS}   & \textbf{0.018}  & 0.023 &     -        &               \textbf{0.2}& 0.024&     -    & \textbf{0.013}  & 0.015          &     - \\ \hline  

\textbf{MAE}    & 0.027 & \textbf{0.024} & 0.025   & 0.028 & 0.026  &   0.026     & 0.016 & \textbf{0.015}&  \textbf{0.015}\\ \hline  

\textbf{RMSE}   & 0.145 & 0.103 & \textbf{0.102}        & 0.25  & \textbf{0.220}  &   0.22      & 0.159   & \textbf{0.138}     &  \textbf{0.138} \\ \hline  

\textbf{mFSS/meFSS}& \textbf{0.34}  & 0.24  & 0.22 & \textbf{0.25}  & 0.2  & 0.18 & \textbf{0.19}  & 0.13  & 0.12 \\ \hline  

\textbf{$\Delta$mFSS} & \textbf{54\%}& 9\%&  - & \textbf{38\%}&  11\%&  - & \textbf{58\%}&  8\%&  - \\ \hline  

\textbf{BIAS} & -0.211& -0.217& \textbf{-0.21}& 0.101&  0.088&  \textbf{0.079}& 0.361&  0.352&  \textbf{0.322}\\ \hline 

\end{tabular}

\end{table}

\subsubsection{Ensemble calibration}\label{sec6_61x}

\begin{figure}[ht]
\centering
\includegraphics[width=1\textwidth]{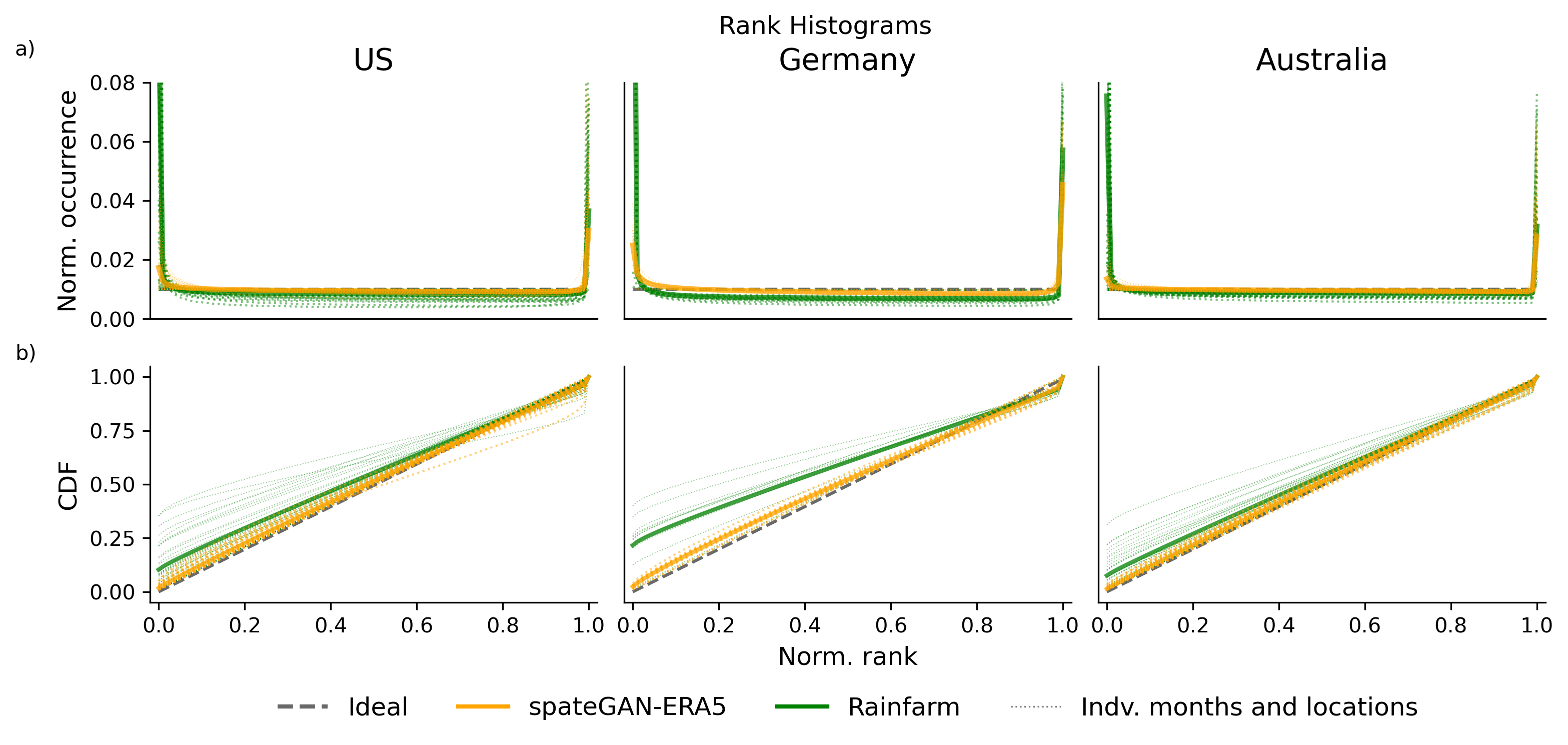}
\caption{Evaluation of the ensemble calibration of the probabilistic models rainFARM and spateGAN-ERA5 for 100 ensemble members of the evaluation dataset from US, Germany and Australia. a) Showing the rank histogram as the per-pixel normalized rank occurrence and b) the cumulative density function of the presented distributions. Dotted lines represent the individual radar regions and months and the solid lines represent the average of the full evaluation datasets.}\label{sup_fig6}
\end{figure}

Fig. \ref{sup_fig6} shows rank histograms and associated cumulative density functions for the US, Germany and Australia, to investigate the calibration of the ensemble of the probabilistic approaches spateGAN-ERA5 and rainFARM. 

\subsubsection{Spatial plausibility - extended analysis}\label{sec6_61}

\begin{figure}[ht]
\centering
\includegraphics[width=1\textwidth]{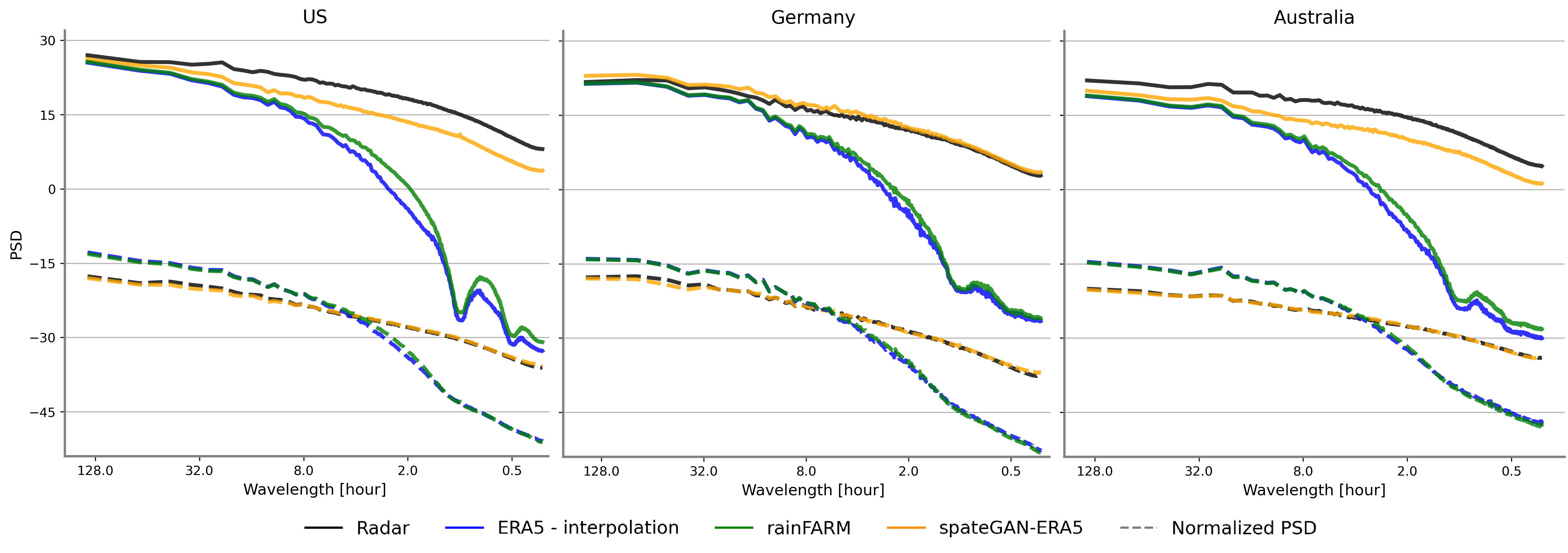}
\caption{Temporal PSD of the full evaluation datasets for the United States, Germany and Australia. }\label{sup_fig7}
\end{figure}

\begin{figure}[ht]
\centering
\includegraphics[width=1\textwidth]{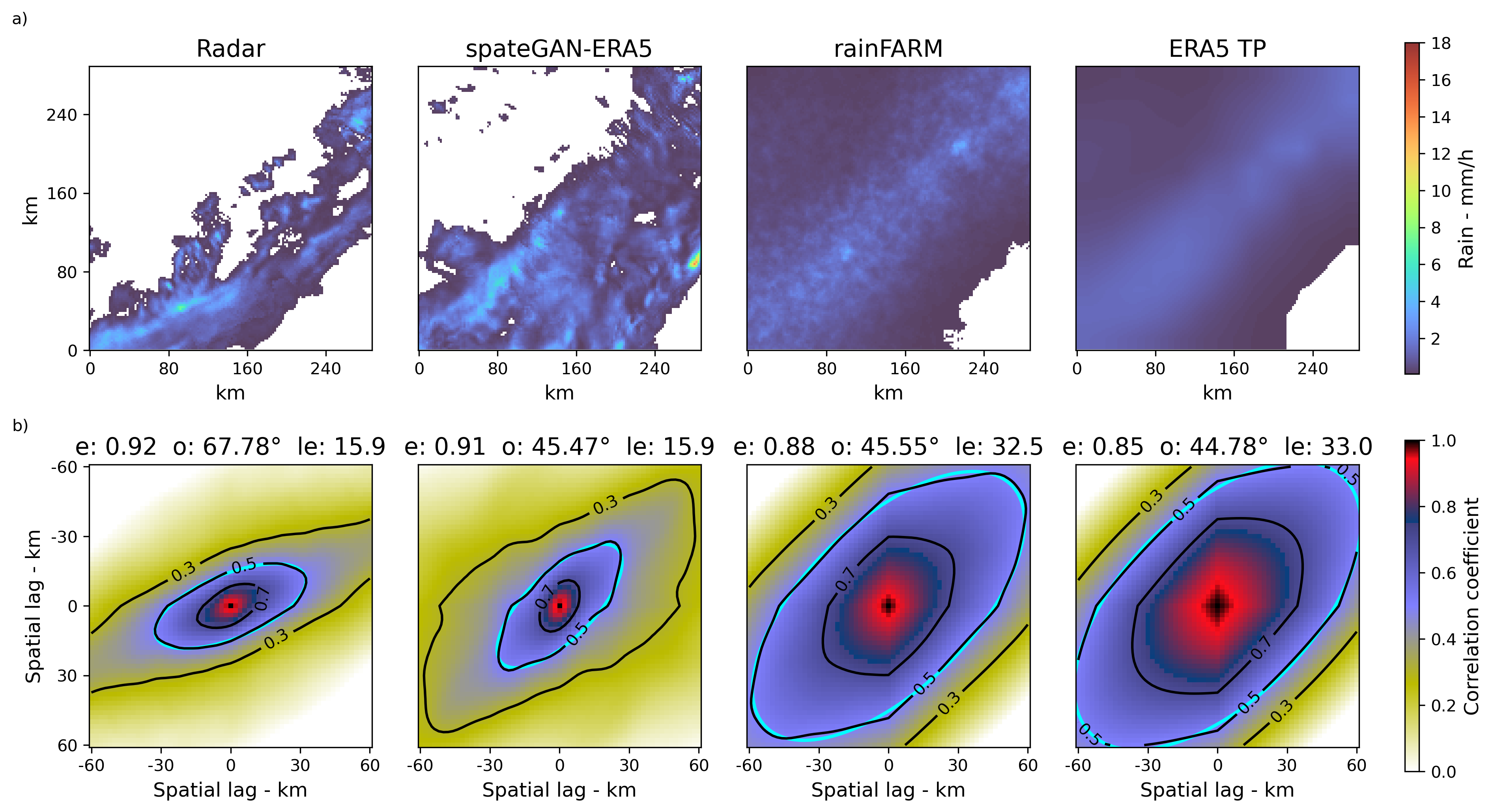}
\caption{a) Exemplary rainfall event in Germany and b) their associated autocorrelation field considered for estimating an ellipse at PCC=0.5 and calculating the eccentricity (e), linear eccentricity (le) and orientation (o).}\label{sup_fig8}
\end{figure}

\begin{figure}[ht]
\centering
\includegraphics[width=1\textwidth]{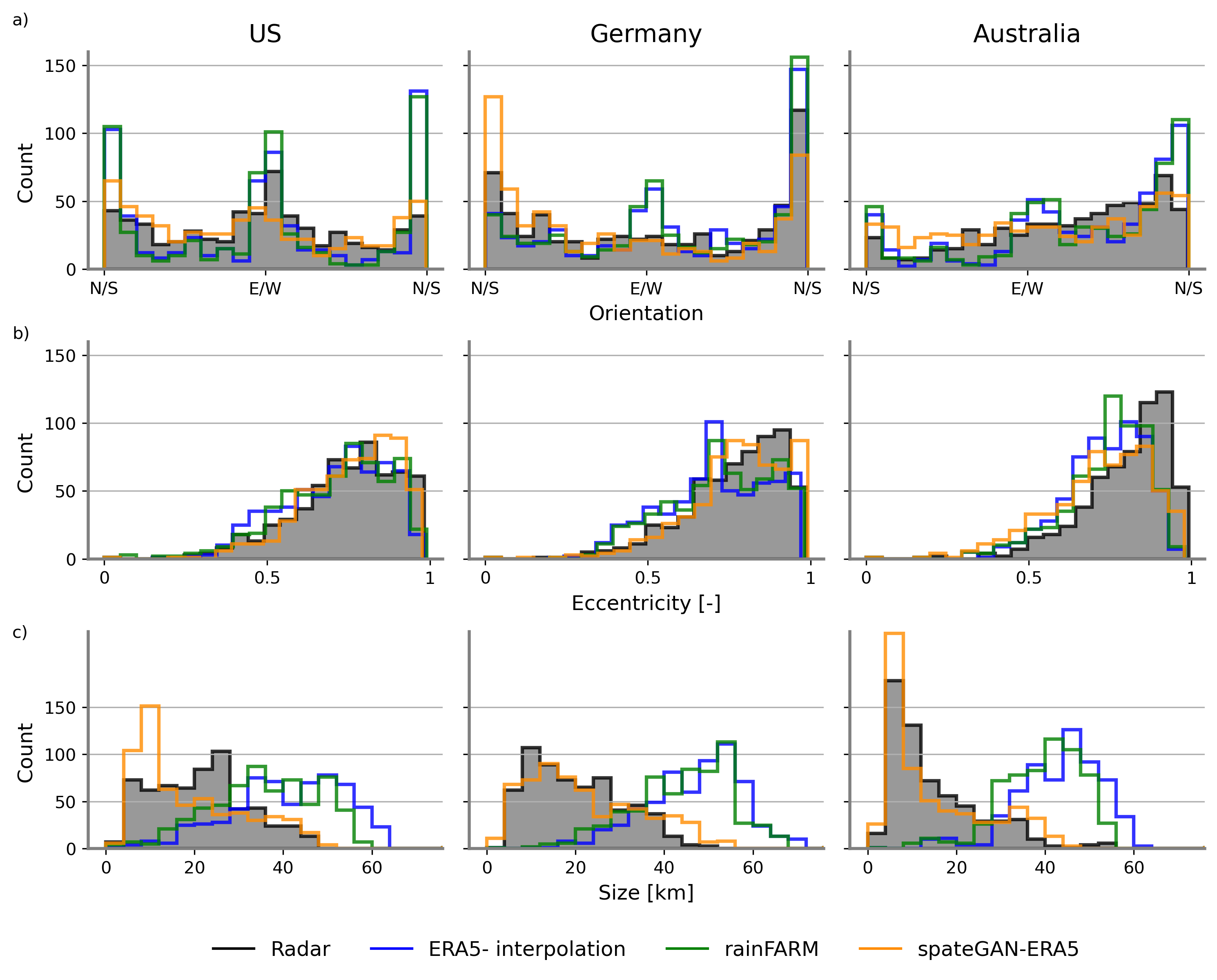}
\caption{Distributions of spatial plausibility scores derived from autocorrelation measurements of  a sub selection of the evaluation datasets, where an increased agreement between ERA5 interpolation and radar observation can be shown (see section \ref{sec6_y}). a) Orientation of the derived 0.5 PCC ellipse (0$°$ and 180 $°$ refers to a North/South orientation), b) distribution of the eccentricity of the ellipse and c) actual size of the ellipse.}\label{sup_fig_ecc}
\end{figure}

Fig. \ref{sup_fig7} shows the temporal PSD of the full evaluation datasets. For Germany, a slight overestimation of the PSD for low and high frequencies is present. For the US and Australia, an overall underestimation can be shown. Shifting the temporal PSD along the y-axis results in a higher agreement, which is also apparent by a good match between the normalized PSD of spateGAN-ERA5 predictions and observations. As described in section \ref{sec5} the deviation of the pure temporal PSD can be explained by the overall observed mean field biases of ERA5 in these regions and its general underestimation of extreme precipitation events \cite{lavers_evaluation_2022}.\\
\\
Fig. \ref{sup_fig8} shows an exemplary rainfall event and its autocorrelation field for the radar observation, spateGAN-ERA5 prediction, rainFARM and ERA5 interpolation, in Germany. The 0.5 PCC counterline is considered for estimating an ellipse, from which the linear eccentricity, eccentricity and orientation can be derived. spateGAN-ERA5 shows the highest similarity to the observed rainfield.
\\\\
Fig. \ref{sup_fig_ecc} shows the distribution of eccentricity, size, and orientation for the subsampled evaluation dataset described in section \ref{sec6_y}. For all three scores and all three regions, a more similar distribution between radar rainfields and spateGAN-ERA5 predictions can be observed. The amount of north/south orientated rainfields ($\approx$0$°$) is reduced for the spateGAN-ERA5 predictions compared to the ERA5 and rainFARM data for the US and Australia. This is in agreement with the radar observations. As can be seen for the training region in Germany, spateGAN-ERA5 does not always predict less rainfields oriented into this direction, compared to ERA5, but also shows an increase for angles between 0-10$°$, which even exceeds the radar distribution. This is another strong indication of sspateGAN-ERA5's generalization capability.


\subsubsection{Additional Case studies}\label{sec6_62}

\begin{figure}[ht]
\centering
\includegraphics[width=1\textwidth]{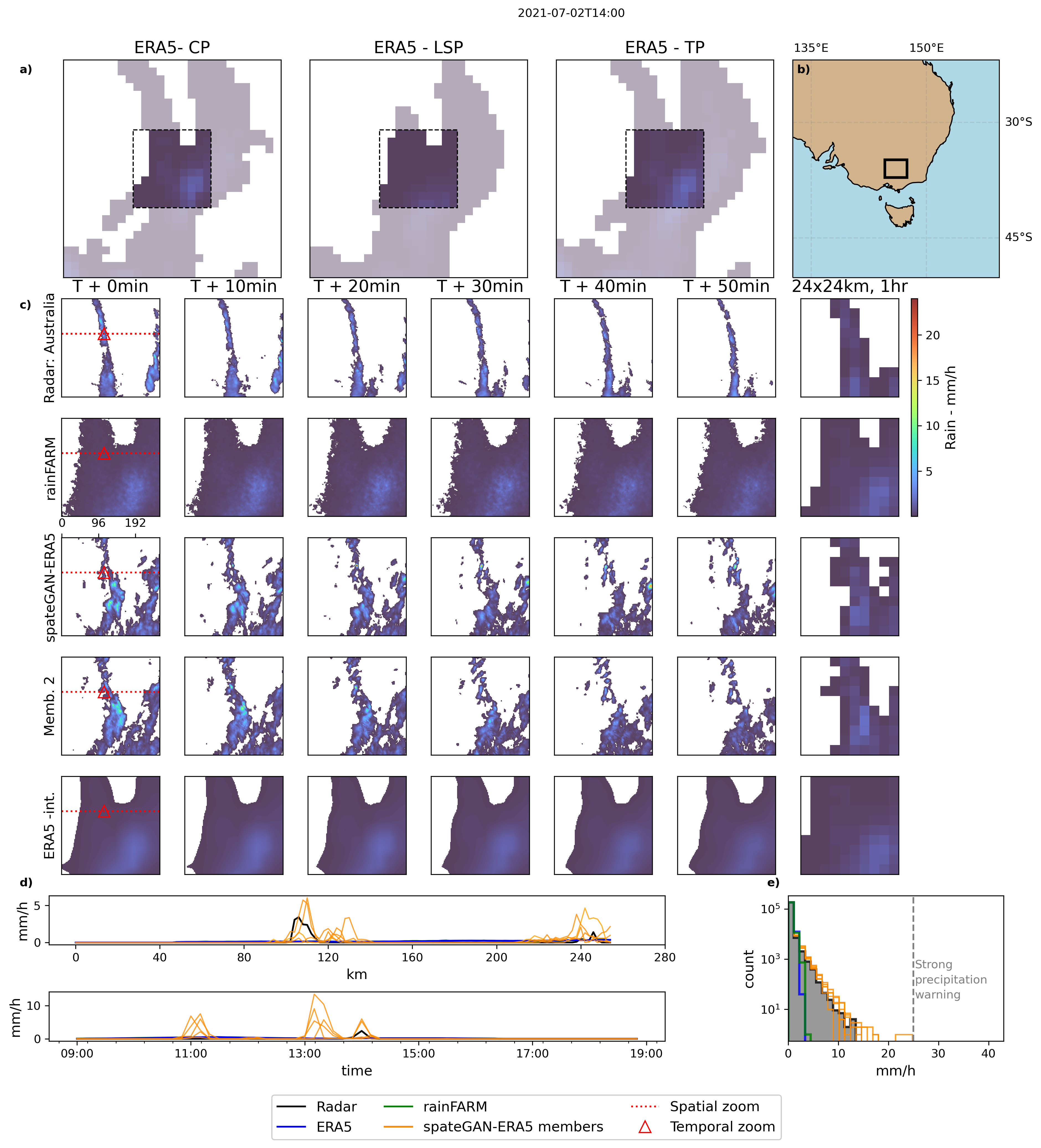}
\caption{Case study event starting on 02.07.21 in Australia. The entire event is shown in the ancillary files, see video V2.  a) Model input patches consisting of larger ERA5 data, i.e. the convective and large-scale precipitation contribution to the total precipitation sum. b) Location of the radar observation. c) Observations, spateGAN-ERA5 predictions and rainFARM downscaling for the target domain in 10-minute increments from t to t+50 min. and as a coarsened version approximating ERA5 resolution. d) 1-D cutouts showing spateGAN-ERA5 ensemble members for a specific pixel along the temporal dimension (top panel) and a horizontal cross-section for one time-step (bottom panel). e) Distribution for temporally aggregated data with 2\,km and 1 hour resolution including maps shown in c) and the previous and following 6 hours. A severe precipitation warning threshold of the German Weather Service is set at 25 mm/h.}\label{sup_fig9}
\end{figure}

\begin{figure}[ht]
\centering
\includegraphics[width=1\textwidth]{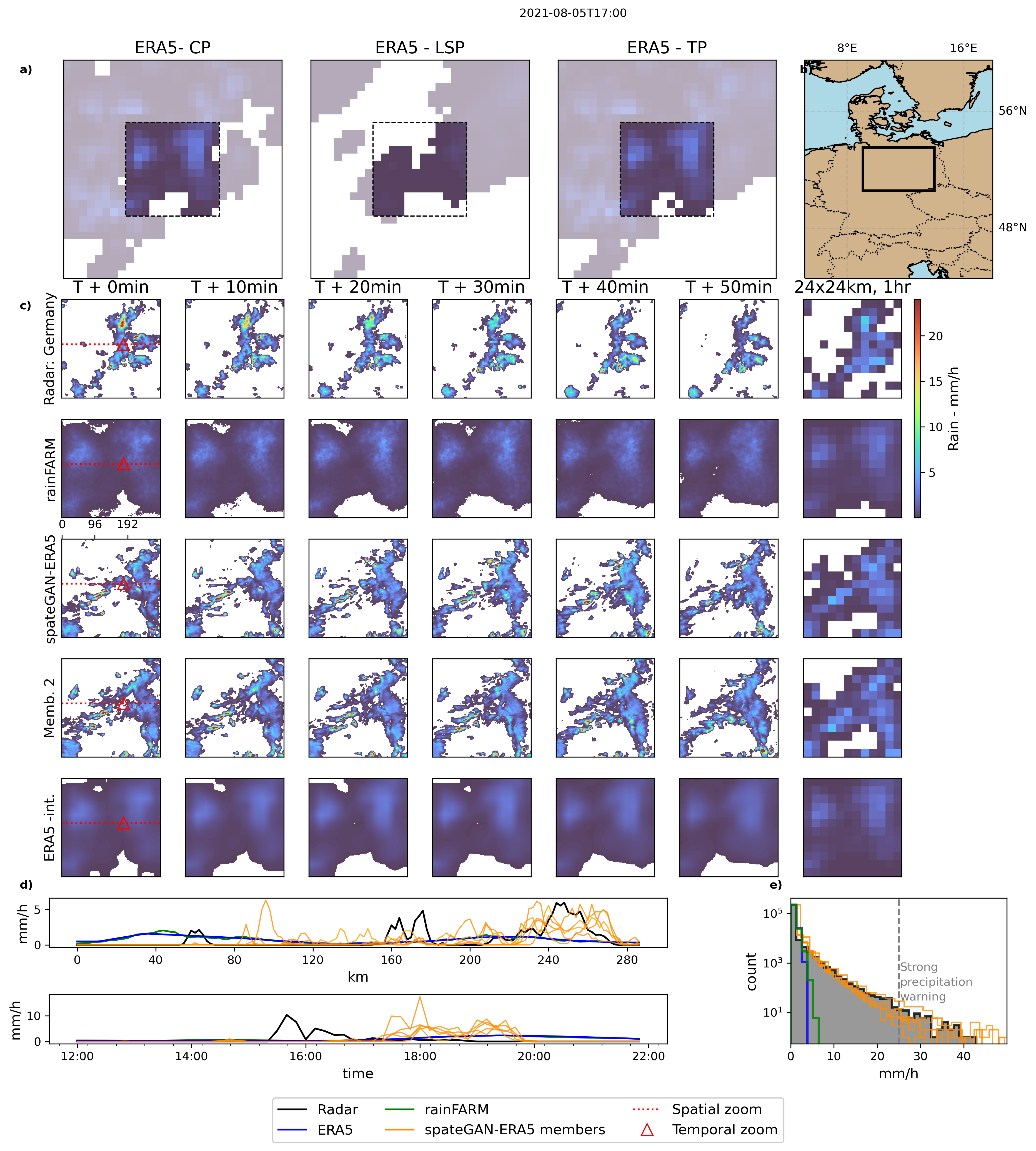}
\caption{Case study as in \ref{sup_fig9} for an event in Germany starting on 05.08.2021. The entire event is shown in the ancillary files, see video V3.}\label{sup_fig10}
\end{figure}

\begin{figure}[ht]
\centering
\includegraphics[width=1\textwidth]{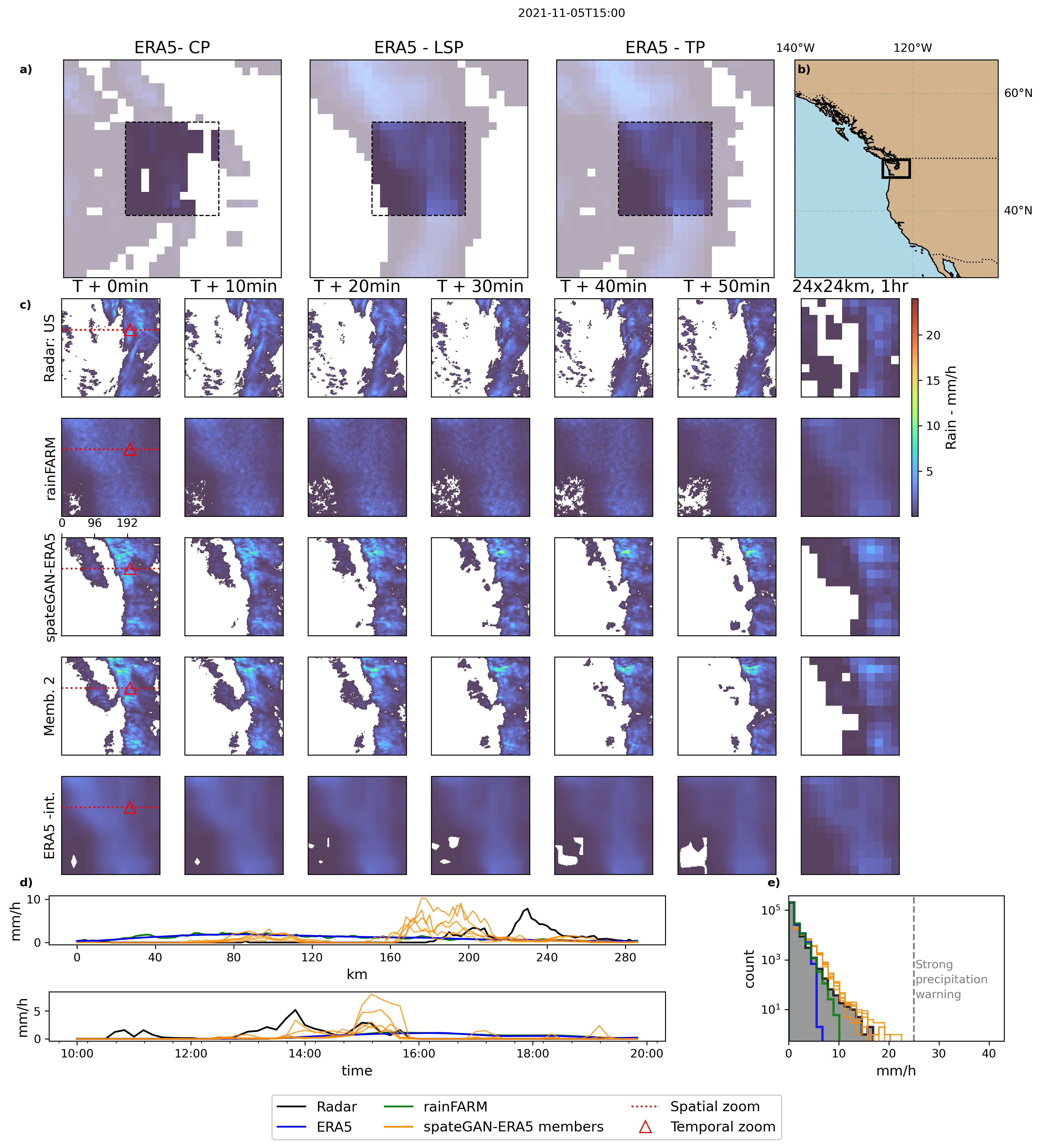}
\caption{Case study as in \ref{sup_fig9} for an event in the US starting on 05.11.2021. The entire event is shown in the ancillary files, see video V4.}\label{sup_fig11}
\end{figure}

In Figs. \ref{sup_fig9}, \ref{sup_fig10}, \ref{sup_fig11} we report three additional case studies for selected events in the US, Australia and Germany.  These events are also shown as videos in the ancillary files.
\\\\
In all cases, ERA5 provides too blurry precipitation fields that do not capture larger rainfall amounts as observed by the radar. RainFARM slightly improves the rainfall distribution but is not capable of disaggregating the individual fields in time and space. SpateGAN-ERA5 predicts rainfields with convective and stratiform character with plausible structure when compared to the reference product. 
In Fig. \ref{sup_fig9} the two easterly moving consecutive rain events are better represented in spateGAN-ERA5 than within the model input. 
In Fig. \ref{sup_fig10} spateGAN-ERA5 is the only method that also resolves the small-scale characteristic of the event while reconstructing hourly rainfall amounts of 40 mm/h. For the third event shown in Fig. \ref{sup_fig11} spateGAN-ERA5 slightly overestimates the medium rainfall intensities while capturing the overall rainfall characteristic with small-scale events in the western part of the image and spatially extended rainfall in the eastern part.
\\\\
All three cases indicate that spateGAN-ERA5 is a powerful spatio-temporal downscaling method that captures not only the spatial characteristics of rainfields like they would be observed by weather radar but also reconstructs the missing tail of the ERA5 precipitation distribution.

\subsubsection{Impacts at ERA5 resolution}\label{sec6_63}

\begin{figure}[ht]
\centering
\includegraphics[width=1\textwidth]{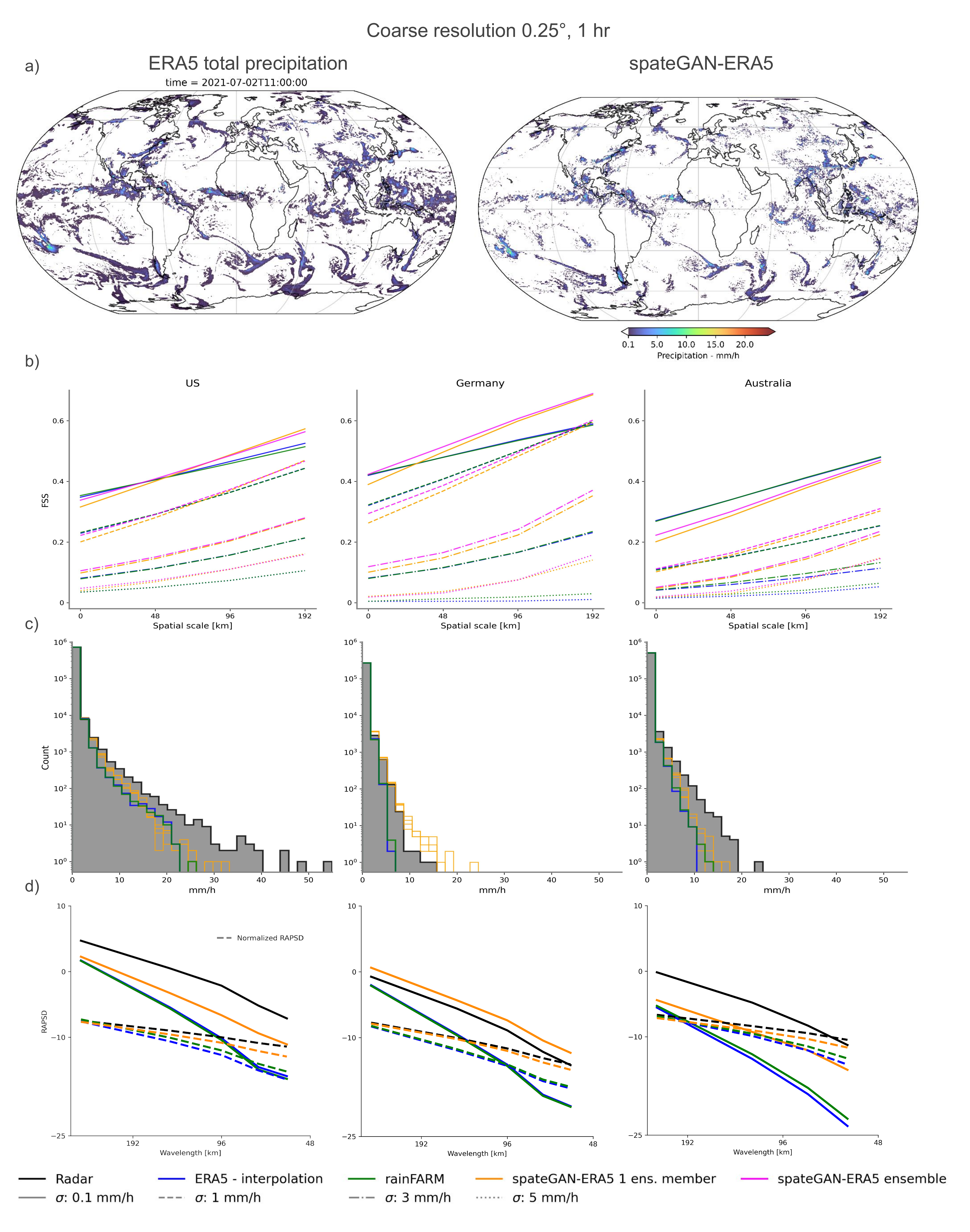}
\caption{a) Shows the global spateGAN-ERA5 predictions coarsened to 1440 x 720 lat-lon grid cells and 1 hour and the associated ERA5 total precipitation estimates for one specific time step. For quantitative evaluation on this resolution, we report b) the FSS, c) the distribution, and d) the RAPSD and normalized RAPSD (dashed lines) of spateGAN-ERA5 and comparison methods of the evaluation datasets for Germany, US and Australia.}\label{sup_fig12}
\end{figure}

Fig. \ref{sup_fig12} shows that even at its native resolution, ERA5 is underestimating larger precipitation intensities when compared to (gauge-adjusted) weather radar observations. This was also reported by the initial ERA5 publication \cite{hersbach_era5_2020}. Furthermore, ERA5 total precipitation is not only limited in its ability to predict extreme values but also strongly underestimates rain rates $>$ 3 mm/h and produces too blurry rainfields as can be seen in the RAPSD for all three considered countries. To demonstrate that spateGAN-ERA5 enhances the statistical properties of ERA5 even at a coarse resolution, we repeated a range of scores using the evaluation datasets coarsened to a 24\,km spatial and hourly temporal resolution.
\\\\
For Germany, the coarsened spateGAN-ERA5 ensemble rainfields show an improved distribution and FSS for the thresholds 0.1, 3 and 5 mm/h on all considered spatial scales, when compared to radar observations. Furthermore, the power spectrum is improved, while slightly overestimating the radars spectrum. Outside of its training region, spateGAN-ERA5 predictions show improved skill, especially for rain rates between 3 mm/h and 15 mm/h. The power spectrum is improved, especially when considering the normalized RAPSD that is less affected by the actual rain rates and the region-dependent ERA5 bias (see Tab. \ref{tab1}). The missing tail of the ERA5 distribution is on this resolution, however, not fully recovered. 
Note that rainFARM stays close to its input data when coarsened. Slight deviations are due to the applied smoothing operation and the considered ensemble.

\subsection{Limitations}\label{sec6_7} 
spateGAN-ERA5 is subject to specific design choices and limitations that need to be considered when applying the method and are, therefore, summarized below.\\
\\
First, spateGAN-ERA5 predictions are constrained to the ERA5 precipitation average of the input patch. It therefore mainly addresses limitations of ERA5 that are related to its coarse resolution. Its performance on a larger scale is therefore bound to the regional and event-specific skill of ERA5. The produced high-resolution maps should be used with prior knowledge about the reanalysis limitations. With prior knowledge, a regional bias correction can be applied to ERA5 precipitation data which will directly translate to the downscaled output of spateGAN-ERA5.\\
\\
Second, spateGAN-ERA5 can not be evaluated on the full globe due to the limited availability of high-quality weather radar data. A sophisticated evaluation was carried out in the US, Australia and Germany, however, most parts of the world could not be covered in this analysis. The presented results are a promising indication that spateGAN-ERA5 provides skilful precipitation estimates for most parts of the world. We therefore encourage its global usage, but also encourage its further region and application-specific validation.\\\\
Third, spateGAN-ERA5 was trained for high-quality gauge-adjusted radar observation in Germany. Our design study explicitly targeted a model that learns how to transform ERA5 precipitation estimates towards the best available high-resolution precipitation product with a high match of observed and modelled rainevents. This study design led, however, to occasional drops in performance for out-of-distribution data with respect to the training domain. Especially extremely strong rain events which rarely occur in Germany can lead to more unrealistic spatial patterns despite a potentially correct estimate of the amplitude.\\
\\
A future step could be to change the study design and include radar observations from different regions of the world in the training data. This, however, increases the heterogeneity in radar data quality and potential mismatches between observations and ERA5 estimates and was, therefore, not considered in this study.

\FloatBarrier


\end{appendices}


\end{document}